\pdfoutput=1
\RequirePackage{ifpdf}
\ifpdf % We are running pdfTeX in pdf mode
\documentclass[pdftex]{sigma}
\else
\documentclass{sigma}
\fi

\usepackage{dsfont}

\numberwithin{equation}{section}

\newtheorem{Theorem}{Theorem}[section]
\newtheorem{Corollary}[Theorem]{Corollary}
\newtheorem{Lemma}[Theorem]{Lemma}
\newtheorem{Proposition}[Theorem]{Proposition}
\newtheorem*{satz}{Theorem}
\newtheorem{scase}[Theorem]{Case}
{\theoremstyle{definition}
\newtheorem{Definition}[Theorem]{Definition}
\newtheorem{Convention}[Theorem]{Convention}
\newtheorem{Example}[Theorem]{Example}
\newtheorem{Remark}[Theorem]{Remark}
}

\begin{document}

\newcommand{\dt}[2]{\frac{\mathrm{d}}{\mathrm{d}#1}\Big|_{#1=#2}}
\newcommand{\dtB}[2]{\frac{\mathrm{d}}{\mathrm{d}#1}\bigg|_{#1=#2}}
\newcommand{\dttB}[2]{{\textstyle\frac{\mathrm{d}}{\mathrm{d}#1}}\big|_{#1=#2}}
\newcommand{\Co}[1]{\alpha_{#1}}
\newcommand{\wt}[1]{\widetilde{#1}}
\newcommand{\Add}[1]{\mathrm{Ad}_{#1}}
\newcommand{\ovl}[1]{\overline{#1}}

\allowdisplaybreaks

\renewcommand{\PaperNumber}{025}

\FirstPageHeading

\ShortArticleName{A Characterization of Invariant Connections}

\ArticleName{A Characterization of Invariant Connections}

\Author{Maximilian HANUSCH}

\AuthorNameForHeading{M.~Hanusch}

\Address{Department of Mathematics, University of Paderborn,\\
Warburger Stra\ss e 100, 33098 Paderborn, Germany}
\Email{\href{mailto:mhanusch@math.upb.de}{mhanusch@math.upb.de}}

\ArticleDates{Received December 09, 2013, in f\/inal form March 10, 2014; Published online March 15, 2014}

\Abstract{Given a~principal f\/ibre bundle with structure group $S$, and a~f\/ibre transitive Lie group~$G$ of automorphisms
thereon, Wang's theorem identif\/ies the invariant connections with certain linear maps $\psi\colon
\mathfrak{g}\rightarrow \mathfrak{s}$.
In the present paper, we prove an extension of this theorem which applies to the general situation where~$G$ acts
non-transitively on the base manifold.
We consider several special cases of the general theorem, including the result of Harnad, Shnider and Vinet which applies
to the situation where~$G$ admits only one orbit type.
Along the way, we give applications to loop quantum gravity.}


\Classification{22F50; 53C05; 53C80; 83C45}

\section{Introduction}

The set of connections on a~principal f\/ibre bundle $(P,\pi,M,S)$ is closed under pullback by automorphisms, and it is
natural to search for connections that do not change under this operation.
Especially, connections invariant under a~Lie group $(G,\Phi)$ of automorphisms are of particular interest as they
ref\/lect the symmetry of the whole group and, for this reason, f\/ind their applications in the symmetry reduction of
(quantum) gauge f\/ield theories~\cite{MathStrucLQG, ChrisSymmLQG, InvConnLQG}. The f\/irst classif\/ication theorem for such
connections was given by Wang~\cite{Wang}, cf.~Case~\ref{th:wang}.
This applies to the case where the induced action\footnote{Each Lie group of automorphisms of a~bundle induces a~smooth
action on the base manifold.} $\varphi$ acts transitively on the base manifold and states that each point in the bundle
gives rise to a~bijection between the set of $\Phi$-invariant connections and certain linear maps $\psi\colon
\mathfrak{g}\rightarrow \mathfrak{s}$.
In~\cite{HarSni} the authors generalize this to the situation where~$\varphi$ admits only one orbit type.
More precisely, they discuss a~variation\footnote{Amongst others, they assume the $\varphi$-stabilizer of $\pi(p_0)$ to
be the same for all $p_0\in P_0$.} of the case where the bundle admits a~submanifold $P_0$ with $\pi(P_0)$ intersecting
each $\varphi$-orbit in a~unique point, see Case~\ref{scase:OneSlice} and Example~\ref{example:SCHSV}.
Here, the $\Phi$-invariant connections are in bijection with such smooth maps $\psi\colon \mathfrak{g}\times
P_0\rightarrow \mathfrak{s}$ for which the restrictions $\psi|_{\mathfrak{g}\times T_{p_0}P_0}$ are linear for all
$p_0\in P_0$, and that fulf\/il additional consistency conditions.

Now, in the general case we consider \emph{$\Phi$-coverings} of $P$.
These are families $\{P_\alpha\}_{\alpha\in I}$ of immer\-sed submanifolds\footnote{For the moment, assume that
$P_\alpha\subseteq P$ is a~subset which, at the same time, is a~manifold such that the inclusion map $\iota_\alpha
\colon P_\alpha \rightarrow P$ is an immersion.
Here, we tacitly identify~$T_{p_\alpha}P_\alpha$ with $\operatorname{\mathrm{im}}[\mathrm{d}_{p_\alpha}\iota_\alpha]$.
Note that we do not require~$P_\alpha$ to be an embedded submanifold of $P$.
For details, see Convention~\ref{conv:Submnfds}.} $P_\alpha$ of $P$ such that each $\varphi$-orbit has non-empty
intersection with $\bigcup_{\alpha\in I}\pi(P_\alpha)$ and for which
\begin{gather*}
T_{p}P = T_{p}P_\alpha +
\mathrm{d}_e\Phi_p(\mathfrak{g}) + Tv_{p}P
\end{gather*}
holds whenever $p\in P_\alpha$ for some $\alpha\in I$.
Here, $Tv_{p}P\subseteq T_pP$ denotes the vertical tangent space at $p\in P$ and $e$ the identity in~$G$.
Observe that the intersection properties of the sets $\pi(P_\alpha)$ with the $\varphi$- orbits in the base manifold
need not to be convenient in any sense.
Indeed, here one might think of situations in which $\varphi$ admits dense orbits, or of the almost-f\/ibre transitive case, cf.
Case~\ref{scase:slicegleichredcluster}.

Let $\Theta\colon (G\times S)\times P\rightarrow P$ be def\/ined by $((g,s),p)\mapsto \Phi(g,p)\cdot s^{-1}$ for $(G,\Phi)$ a Lie group of automorphisms of $(P,\pi,M,S)$. Then, the main
result of the present paper can be stated as follows:

\begin{satz}
Each $\Phi$-covering $\{P_\alpha\}_{\alpha\in I}$ of $P$ gives rise to a~bijection between the $\Phi$-invariant connections on $P$ and the families $\{\psi_\alpha\}_{\alpha\in I}$ of smooth maps $\psi_\alpha\colon \mathfrak{g}\times TP_\alpha
\rightarrow \mathfrak{s}$ for which ${\psi_\alpha}|_{\mathfrak{g}\times T_{p_\alpha} P_\alpha}$ is linear for all
$p_\alpha\in P_\alpha$, and that fulfil the following two $($generalized Wang$)$ conditions:
\begin{itemize}\itemsep=0pt
\item
$\wt{g}(p_\beta) + \vec{w}_{p_\beta}-\wt{s}(p_\beta)={\rm d}L_q\vec{w}_{p_\alpha}
\quad
\Longrightarrow
\quad
\psi_\beta(\vec{g},\vec{w}_{p_\beta})-\vec{s}=\rho(q)\circ\psi_\alpha\big(\vec{0}_{\mathfrak{g}},\vec{w}_{p_\alpha}\big)$,
\item
$\psi_\beta\big(\Add{q}(\vec{g}),\vec{0}_{p_\beta}\big)=\rho(q)\circ
\psi_\alpha\big(\vec{g},\vec{0}_{p_\alpha}\big)$
\end{itemize}
with $\rho(q):=\Add{s}$ and $\Add{q}(\vec{g}):=\Add{g}(\vec{g})$ for $q=(g,s)\in Q$.

 Here, $\wt{g}$ and $\wt{s}$ denote the fundamental vector fields that correspond to the elements $\vec{g}\in \mathfrak{g}$ and
$\vec{s}\in \mathfrak{s}$, respectively; and of course we have $\vec{0}_{p_\alpha},\vec{w}_{p_\alpha}\in T_{p_\alpha}P_\alpha$, $\vec{0}_{p_\beta},\vec{w}_{p_\beta}\in T_{p_\beta}P_\beta$ as well as $p_\beta=q\cdot p_\alpha$ for $p_\alpha\in P_\alpha, p_\beta\in P_\beta$.
\end{satz}

Using this theorem, the calculation of invariant connections reduces to identifying a~$\Phi$-covering which makes the
above conditions as easy as possible.
Here, one basically has to f\/ind the balance between quantity and complexity of these conditions.
Of course, the more submanifolds there are, the more conditions we have, so that usually it is convenient to use as few
of them as possible.
For instance, in the situation where $\varphi$ is transitive, it suggests itself to choose a~$\Phi$-covering that
consists of one single point; which, in turn, has to be chosen appropriately.
Also if there is some $m\in M$ contained in the closure of each $\varphi$-orbit, one single submanifold is suf\/f\/icient,
see Case~\ref{scase:slicegleichredcluster} and Example~\ref{ex:Bruhat}.
The same example also shows that sometimes pointwise\footnote{Here, pointwise means to consider such elements $q\in G\times S$
that are contained in the $\Theta$-stabilizer of some f\/ixed $p_\alpha\in P_\alpha$ for $\alpha\in I$.} evaluation of the
above conditions proves non-existence of $\Phi$-invariant connections.

\looseness=-1
In any case, one can use the inverse function theorem to construct a~$\Phi$-covering $\{P_\alpha\}_{\alpha\in I}$ of~$P$
such that the submanifolds~$P_\alpha$ have minimal dimension in a~certain sense, see Lemma~\ref{lemma:mindimslice} and Co\-rol\-la\-ry~\ref{cor:reductions}.
This reproduces the description of connections by means of local 1-forms on~$M$ provided that~$G$ acts trivially or,
more generally, via gauge transformations on~$P$, see Case~\ref{scase:GaugeTransf}.

Finally, since orbit structures can depend very sensitively on the action or the group, one cannot expect to have
a~general concept for f\/inding the $\Phi$-covering optimal for calculations.
Indeed, sometimes these calculations become easier if one uses coverings that seem less optimal at a~f\/irst sight (as, e.g., if they have no minimal dimension, cf.\ calculations in Appendix~\ref{subsec:IsotrConn}).

The present paper is organized as follows:
%\begin{itemize}\itemsep=0pt
%\item
In Section~\ref{sec:prep}, we f\/ix the notations.
%\item
In Section~\ref{sec:phiCoverings}, we introduce the notion of a~$\Phi$-covering, the central object of this paper.
%\item
In Section~\ref{sec:mainth}, we prove the main theorem and deduce a~slightly more general version of the result from~\cite{HarSni}.
%\item
In Section~\ref{sec:PartCases}, we show how to construct $\Phi$-coverings to be used in special situations.
In particular, we consider the (almost) f\/ibre transitive case, trivial principal f\/ibre bundles and Lie groups of gauge
transformations.
Along the way, we give applications to loop quantum gravity.
%\end{itemize}

\section{Preliminaries}\label{sec:prep}

We start with f\/ixing the notations.

\subsection{Notations}\label{sec:notations}

Manifolds are always assumed to be smooth.
If $M$, $N$ are manifolds and $f\colon M\rightarrow N$ is a~smooth map, then $\mathrm{d} f\colon TM\rightarrow TN$ denotes
the dif\/ferential map between their tangent manifolds.
The map $f$ is said to be an immersion if\/f for each $x\in M$ the restriction $\mathrm{d}_xf:=\mathrm{d} f|_{T_xM}\colon
T_xM\rightarrow T_{f(x)}N$ is injective.

Let $V$ be a~f\/inite dimensional vector space.
A $V$-valued 1-form $\omega$ on the manifold $N$ is a~smooth map $\omega\colon TN\rightarrow V$ whose restriction
$\omega_y:=\omega|_{T_yN}$ is linear for all $y\in N$.
The pullback of $\omega$ by $f$ is the $V$-valued 1-form $f^{*}\omega\colon TM\rightarrow V$, $\vec{v}_x\rightarrow
\omega_{f(x)}(\mathrm{d}_xf(\vec{v}_x))$.

Let~$G$ be a~Lie group with Lie algebra $\mathfrak{g}$.
For $g\in G$, we def\/ine the corresponding conjugation map by $\Co{g}\colon G\rightarrow G$, $h\mapsto g h g^{-1}$.
Its dif\/ferential $\mathrm{d}_e\Co{g}\colon \mathfrak{g}\rightarrow \mathfrak{g}$ at the unit element $e\in G$ is denoted
by $\Add{g}$ in the following.

Let $\Psi$ be a~(left) action of the Lie group~$G$ on the manifold~$M$.
For $g\in G$ and $x\in M$, we define $\Psi_g\colon M\rightarrow M$, $\Psi_g\colon y\mapsto \Psi(g,y)$ and $\Psi_x\colon G\rightarrow M$, $h\mapsto\Psi(h,x)$, respectively. If it is clear which action is meant, we will often write $L_g$ instead of $\Psi_g$ as well as $g\cdot y$ or $g y$ instead of $\Psi_g(y)$. For $\vec{g}\in \mathfrak{g}$ and $x\in M$, the map
\begin{gather*}
\wt{g}(x):=\dttB{t}{0} \Psi_x(\exp(t\vec{g}\hspace{1pt}))
\end{gather*}
is called the
\emph{fundamental vector field of~$\vec{g}$}.
The Lie subgroup $G_x:=\left\{g\in G\: \big| \: g\cdot x=x\right\}$ is called the \emph{stabilizer} of $x\in M$
(w.r.t.~$\Psi$), and its Lie algebra $\mathfrak{g}_x$ equals $\ker[\mathrm{d}_x\Psi]$, see e.g.~\cite{DuisKolk}.
The \emph{orbit} of $x$ under~$G$ is the set $Gx:=\operatorname{\mathrm{im}}[\Psi_x]$. $\Psi$ is said to be
\emph{transitive} if\/f $Gx=M$ holds for one (and then each) $x\in M$.
Analogous conventions we also use for right actions.

\subsection{Invariant connections}\label{subsec:InvConn}

Let $\pi\colon P\rightarrow M$ be a~smooth map between manifolds $P$ and~$M$, and denote by
$F_x:=\pi^{-1}(x)\subseteq P$ the f\/ibre over $x\in M$ in $P$.
Moreover, let $S$ be a a~Lie group that acts via $R\colon P\times S\rightarrow P$ from the right on $P$.
If there is an open covering $\{U_\alpha\}_{\alpha\in I}$ of~$M$ and a~family $\{\phi_\alpha\}_{\alpha\in I}$ of
dif\/feomorphisms $\phi_\alpha\colon \pi^{-1}(U_\alpha)\rightarrow U_\alpha\times S$ with
\begin{gather}
\label{eq:bundlemaps}
\phi_\alpha(p\cdot s)=\big(\pi(p),[\mathrm{pr}_2\circ\phi_\alpha](p)\cdot s\big)
\qquad\forall\, p\in \pi^{-1}(U_\alpha),\qquad\forall\, s\in S,
\end{gather}
then $(P,\pi,M,S)$ is called \emph{principal fibre bundle} with total space $P$,
projection map $\pi$, base manifold~$M$ and structure group $S$.
Here, $\mathrm{pr}_2$
denotes the projection onto the second factor.
It follows from~\eqref{eq:bundlemaps} that $\pi$ is surjective, and that:
\begin{itemize}\itemsep=0pt
\item
$R_s(F_x)\subseteq F_x$ for all $x\in M$ and all $s\in S$,
\item
for each $x\in M$ the map $R_x\colon F_x \times S\rightarrow F_x$, $(p,s)\mapsto p\cdot s$ is transitive and free.
\end{itemize}
The subspace $Tv_pP:=\ker[d_p\pi]\subseteq T_pP$ is called \emph{vertical tangent
space} at $p\in P$ and
\begin{gather*}
\wt{s}(p):=\dttB{t}{0}\: p\cdot \exp(t\vec{s}\hspace{1pt})\in Tv_pP
\qquad\forall\, p\in P,
\end{gather*}
denotes the fundamental vector f\/ield of $\vec{s}$ w.r.t.~the right action of $S$ on
$P$.
The map $\mathfrak{s}\ni\vec{s}\rightarrow \wt{s}(p)\in Tv_pP$ is a~vector space isomorphism for all $p\in P$.

Complementary to that, a~\emph{connection} $\omega$ is an $\mathfrak{s}$-valued 1-form on $P$ with \begin{itemize}\itemsep=0pt
\item
$R_s^{*}\omega= \Add{s^{-1}}\circ \omega\qquad \forall\,s\in S$,
\item
$\omega_p(\wt{s}(p))=\vec{s}\hspace{44pt}\forall\,\vec{s}\in \mathfrak{s}$.
\end{itemize}
The subspace $Th_pP:=\ker[\omega_p]\subseteq T_pP$ is called the \emph{horizontal
tangent space} at $p$ (w.r.t.~$\omega$).
We have $\mathrm{d} R_s(Th_pP)=Th_{p\cdot s}P$ for all $s\in S$, and one can show that $T_pP= Tv_pP\oplus Th_pP$ holds for all
$p\in P$.

A dif\/feomorphism $\kappa\colon P\rightarrow P$ is said to be an \emph{automorphism} if\/f $\kappa(p\cdot s)=\kappa(p)\cdot
s$ holds for all $p\in P$ and all $s\in S$.
It is straightforward to see that an $\mathfrak{s}$-valued 1-form $\omega$ on $P$ is a~connection if\/f this is true for
the pullback $\kappa^{*}\omega$.
A \emph{Lie group of automorphisms $(G,\Phi)$ of $P$} is a~Lie group~$G$ together with a~left action $\Phi$ of~$G$ on
$P$ such that the map $\Phi_g$ is an automorphism for each $g\in G$.
This is equivalent to say that $\Phi(g,p\cdot s)=\Phi(g,p)\cdot s$ holds for all $p\in P$, $g\in G$ and all $s\in S$.
In this situation, we will often write $gps$ instead of $(g\cdot p)\cdot s=g\cdot(p\cdot s)$.
Each such a~left action $\Phi$ gives rise to two further actions: \begin{itemize}\itemsep=0pt
\item
The induced action $\varphi$ is def\/ined by
\begin{gather}
\label{eq:INDA}
\begin{split}
&\varphi\colon\quad G\times M  \rightarrow  M,
\\
& \phantom{\varphi\colon\quad{}}
(g,m) \mapsto  (\pi\circ\Phi)(g, p_m),
\end{split}
\end{gather}
where $p_m\in \pi^{-1}(m)$ is arbitrary.
$\Phi$ is called \emph{fibre transitive} if\/f $\varphi$ is transitive.
\item
We equip $Q=G\times S$ with the canonical Lie group structure and def\/ine~\cite{Wang}
\begin{gather}
\label{eq:THETA}
\begin{split}
&\Theta\colon\quad Q \times P \rightarrow  P,
\\
&\phantom{\Theta\colon\quad{}}
((g,s),p) \mapsto  \Phi\left(g, p\cdot s^{-1}\right).
\end{split}
\end{gather}
\end{itemize}
A connection $\omega$ is said to be $\Phi$-invariant if\/f $\Phi_g^{*}\omega=\omega$ holds for all
$g\in G$.
This is equivalent to require that for each $p\in P$ and $g\in G$ the dif\/ferential $\mathrm{d}_pL_g$ induces an
isomorphism between the horizontal tangent spaces $Th_pP$ and $Th_{gp}P$.\footnote{In literature sometimes the latter
condition is used to def\/ine $\Phi$-invariance of connections.}

We conclude this subsection with the following straightforward facts, see also~\cite{Wang}:
\begin{itemize}\itemsep=0pt
\item
Consider the representation $\rho\colon Q \rightarrow \mathsf{Aut}(\mathfrak{s})$, $(g,s)\mapsto \Add{s}$.
Then, it is straightforward to see that each $\Phi$-invariant connection $\omega$ is of type $\rho$, i.e., $\omega$ is an
$\mathfrak{s}$-valued 1-form on $P$ with $L_{q}^{*}\omega=\rho(q)\circ \omega$ for all $q\in Q$.
\item
An $\mathfrak{s}$-valued 1-form $\omega$ on $P$ with $\omega(\wt{s}(p))=\vec{s}$ for all $\vec{s}\in
\mathfrak{s}$ is a~$\Phi$-invariant connection if\/f it is of type $\rho$.
\item
Let $Q_{p}$ denote the stabilizer of $p\in P$ w.r.t.~$\Theta$, and $G_{\pi(p)}$ the stabilizer of
$\pi(p)$ w.r.t.~$\varphi$.
Then, $G_{\pi(p)}=\left\{h\in G\: | \:L_h\colon F_{\pi(p)}\rightarrow F_{\pi(p)} \right\}$, and we obtain a~Lie group homomorphism
\begin{gather*}
 \phi_p\colon G_{\pi(p)}\rightarrow S\quad\text{by requiring that}\quad \Phi(h,p)=p\cdot\phi_p(h)\quad\text{for all}\quad h\in
G_{\pi(p)}.
\end{gather*}
If~$\mathfrak{q}_p$ and~$\mathfrak{g}_{\pi(p)}$ denote the Lie algebras of $Q_p$ and $G_{\pi(p)}$, respectively, then
\begin{gather}
\label{eq:staoQ}
Q_p=\{(h,\phi_p(h))\:|\:h\in G_{\pi(p)}\}
\qquad
\text{and}
\qquad
\mathfrak{q}_p=\big\{\big(\vec{h},
\mathrm{d}_e\phi_p\big(\vec{h}\hspace{1pt}\big)\big)\:\big|\:\vec{h}\in
\mathfrak{g}_{\pi(p)}\big\}.
\end{gather}
\end{itemize}

\section[$\Phi$-coverings]{$\boldsymbol{\Phi}$-coverings}\label{sec:phiCoverings}

We start this section with some facts and conventions concerning submanifolds.
Then, we provide the def\/inition of a~$\Phi$-covering and discuss some its properties.
\begin{Convention}
\label{conv:Submnfds}
Let~$M$ be a~manifold.
\begin{enumerate}\itemsep=0pt
\item[$1.$]
A pair $(N,\tau_N)$ consisting of a~manifold $N$ and an injective immersion $\tau\colon N\rightarrow M$
is called submanifold of~$M$.
\item[$2.$]
If $(N,\tau_N)$ is a~submanifold of~$M$, we tacitly identify $N$ and $TN$ with their images
\mbox{$\tau_N(N) \subseteq  M$} and $\mathrm{d}\tau_N(TN)\subseteq TM$, respectively.
In particular, this means that:{\samepage
\begin{itemize}\itemsep=0pt
\item
If $M'$ is a~manifold and $\kappa\colon M\rightarrow M'$ a~smooth map, then for $x\in N$ and
$\vec{v}\in TN$ we write $\kappa(x)$ and $\mathrm{d}\kappa(\vec{v})$ instead of $\kappa(\tau_N(x))$ and
$\mathrm{d}\kappa( \mathrm{d}\tau(\vec{v}))$, respectively.
\item
If $\Psi\colon G\times M\rightarrow M$ is a~left action of the Lie group~$G$ and $(H,\tau_H)$ a~submanifold of~$G$, the restriction of $\Psi$ to $H\times N$ is def\/ined by
\begin{gather*}
\Psi|_{H\times N}(h,x):= \Psi(\tau_H(h),\tau_N(x))
\qquad\forall\, (h,x)\in H\times N.
\end{gather*}
\item
If $\omega\colon TM\rightarrow V$ is a~$V$-valued 1-form on~$M$, we let
\begin{gather*}
(\Psi^{*}\omega)|_{TG\times TN}(\vec{m},\vec{v}):=(\Psi^{*}\omega)(\vec{m},\mathrm{d}\tau(\vec{v}))
\qquad\forall\, (\vec{m},\vec{v})\in TG\times TN.
\end{gather*}
\item
We will not explicitly refer to the maps $\tau_N$ and $\tau_H$ in the following.
\end{itemize}}
\item[$3.$]
Open subsets $U\subseteq M$ are equipped with the canonical manifold structure making the inclusion map an
embedding.
\item[$4.$]
If $L$ is a~submanifold of $N$, and $N$ is a~submanifold of~$M$, we consider $L$ as a~submanifold of~$M$
in the canonical way.
\end{enumerate}
\end{Convention}

\begin{Definition}
A submanifold $N\subseteq M$ is called $\Psi$-patch if\/f for each $x\in N$ we find an open neighbourhood $N'\subseteq N$
of $x$ and a~submanifold $H$ of~$G$ through $e$, such that the restriction $\Psi|_{H\times N'}$ is a~dif\/feomorphism to an
open subset $U\subseteq M$.
\end{Definition}

\begin{Remark}\label{rem:patch}\quad
\begin{enumerate}\itemsep=0pt
\item
It follows from the inverse function theorem and\footnote{The sum is not necessarily direct.}
\begin{gather*}
\mathrm{d}_{(e,x)}\Psi(\mathfrak{g}\times T_xN)= \mathrm{d}_e\Psi_x(\mathfrak{g}) + \mathrm{d}_x\Psi_e(T_xN)=
\mathrm{d}_e\Psi_x(\mathfrak{g})+T_xN
\qquad\forall\, x\in N
\end{gather*}
that $N$ is a~$\Psi$-patch if\/f $T_xM= \mathrm{d}_e\Psi_x(\mathfrak{g})+T_xN$ holds for all $x\in N$.\footnote{In fact, let $V\subseteq \mathrm{d}_e\Psi_x(\mathfrak{g})$ be an algebraic complement of $T_xN$ in $T_xM$ and $V'\subseteq
\mathfrak{g}$ a~linear subspace with $\dim[V']=\dim[V]$ and $\mathrm{d}_e\Psi_x(V')=V$.
Then, we f\/ind a~submanifold $H$ of~$G$ through $e$ with $T_eH=V'$, so that $\mathrm{d}_{(e,x)}\Psi\colon T_eH\times T_xN
\rightarrow T_xM$ is bijective.}
\item
Open subsets $U\subseteq M$ are always $\Psi$-patches.
They are of maximal dimension, which, for instance, is necessary if there is a~point in $U$ whose stabilizer equals~$G$,
see Lemma~\ref{lemma:mindimslice}.1.
\item
We allow zero-dimensional patches, i.e., $N=\{x\}$ for some $x\in M$.
Necessarily, then we have $\mathrm{d}_e\Psi_x(\mathfrak{g})=T_xM$ as well as $\Psi|_{H\times N}=\Psi_x|_{H}$ for each submanifold $H$
of~$G$.
\end{enumerate}
\end{Remark}

The second part of the following elementary lemma equals Lemma~2.1.1 in~\cite{DuisKolk}.
\begin{Lemma}\label{lemma:mindimslice}
Let $(G,\Psi)$ be a~Lie group that acts on the manifold~$M$, and let $x\in M$.
\begin{enumerate}\itemsep=0pt%\label{item:s}
\item[$1.$]
If $N$ is a~$\Psi$-patch with $x\in N$, then $\dim[N]\geq \dim[M]-\dim[G]+\dim[G_x]$.
\item[$2.$]
Let $V$ and $W$ be algebraic complements of $\mathrm{d}_e\Psi_x(\mathfrak{g})$ in $T_xM$ and of $\mathfrak{g}_x$ in
$\mathfrak{g}$, respectively.
Then there are submanifolds $N$ of~$M$ through $x$ and $H$ of~$G$ through $e$ such that $T_xN=V$, $T_eH=W$.
In particular, $N$ is a~$\Psi$-patch and $\dim[N]=\dim[M]-\dim[G]+\dim[G_x]$.
\end{enumerate}
\end{Lemma}

\begin{proof}
1.~By Remark~\ref{rem:patch}.1 and since $\ker[\mathrm{d}_e\Psi_x]=\mathfrak{g}_x$, we have
\begin{gather}
\label{eq:leq}
\dim[M]\leq \dim[\mathrm{d}_e\Psi_x(\mathfrak{g})]+\dim[T_xN]=\dim[G]-\dim[G_x]+\dim[N].
\end{gather}

2.~Of course, we f\/ind submanifolds $N'$ of~$M$ through $x$ and $H'$ of~$G$ through $e$ such that $T_xN'=V$ and $T_eH'=W$.
So, if $\vec{g}\in \mathfrak{g}$ and $\vec{v}_x\in T_xN'$, then
$0=\mathrm{d}_{(e,x)}\Psi(\vec{g},\vec{v}_x)=\mathrm{d}_e\Psi_x(\vec{g})+\vec{v}_x$ implies
$\mathrm{d}_e\Psi_x(\vec{g})=0$ and $\vec{v}_x=0$.
Hence, $\vec{g}\in \ker[\mathrm{d}_e\Psi_x]=\mathfrak{g}_x$, so that\footnote{Recall that
$\mathrm{d}_{(e,x)}\Psi|_{T_eH'\times T_eN'}\colon \big(\vec{h},\vec{v}_x\big)\mapsto
\mathrm{d}_{(e,x)}\Psi\big(\mathrm{d}_e\tau_H(\vec{h}),\mathrm{d}_x\tau_N(\vec{v}_x)\big)$.}
$\mathrm{d}_{(e,x)}\Psi|_{T_eH'\times T_eN'}$ is injective.
It is immediate from the def\/initions that this map is surjective, so that by the inverse function theorem we f\/ind open
neighbourhoods $N\subseteq N'$ of $x$ and $H\subseteq G$ of $e$ such that $\Psi|_{H\times N}$ is a~dif\/feomorphism to an
open subset $U\subseteq M$.
Then $N$ is a~$\Psi$-patch, and since in~\eqref{eq:leq} equality holds, also the last claim is clear.
\end{proof}

\begin{Definition}\looseness=1%\label{def:pmappe}
Let $(G,\Phi)$ be a~Lie group of automorphisms of the principal f\/ibre bundle~$P$, and recall the actions $\varphi$ and
$\Theta$ def\/ined by~\eqref{eq:INDA} and~\eqref{eq:THETA}, respectively.
A family of $\Theta$-patches $\{P_\alpha\}_{\alpha\in I}$ is said to be a~$\Phi$-covering of $P$ if\/f each
$\varphi$-orbit intersects at least one of the sets~$\pi(P_\alpha)$.
\end{Definition}

\begin{Remark}\label{bem:Psliceeigensch}{\samepage
\quad
\begin{enumerate}\itemsep=0pt
\item
If $O\subseteq P$ is a~$\Theta$-patch, Lemma~\ref{lemma:mindimslice}.1 and~\eqref{eq:staoQ} yield
\begin{gather*}
\dim[O]\geq \dim[P]-\dim[Q]+\dim[Q_p] \stackrel{\eqref{eq:staoQ}}{=}\dim[M]-\dim[G]+\dim[G_{\pi(p)}].
\end{gather*}
\item
It follows from Remark~\ref{rem:patch}.1 and
$\mathrm{d}_e\Theta_p(\mathfrak{q})=\mathrm{d}_e\Phi_p(\mathfrak{g}) + Tv_pP$ that $O$ is a~$\Theta$-patch if\/f
\begin{gather}
\label{eq:transv}
T_pP=T_pO+ \mathrm{d}_e\Phi_p(\mathfrak{g}) + Tv_pP
\qquad\forall\, p\in O.
\end{gather}
As a consequence,
\begingroup
\setlength{\leftmarginii}{19pt}
\begin{itemize}\itemsep=0pt
\item
each $\Phi$-patch is a~$\Theta$-patch,
\item
$P$ is always a~$\Phi$-covering by itself. Moreover, if $P=M\times S$ is trivial, then $M\times \{e\}$ is a~$\Phi$-covering.
\end{itemize}
\item
If $N$ is a~$\varphi$-patch and $s_0\colon N \rightarrow P$ a~smooth section (i.e., $\pi\circ s_0
=\mathrm{id}_N$), then $s_0(N)$ is a~$\Theta$-patch by Lemma~\ref{lemma:suralpha}.2.

Conversely, if $N\subseteq M$ is a submanifold such that $s_0(N)$ is a~$\Theta$-patch for $s_0$ as above, then~$N$ is a $\varphi$-patch. In fact, applying $\mathrm{d}\pi$ to \eqref{eq:transv}, this is immediate from Remark \ref{rem:patch}.1 and the definition of $\varphi$.
\end{enumerate}
\endgroup
}
\end{Remark}

\begin{Lemma}\label{lemma:suralpha}
Let $(G,\Phi)$ be a~Lie group of automorphisms of the principal bundle $(P,\pi,M,S)$.
\begin{enumerate}\itemsep=0pt
\item[$1.$]
If $O\subseteq P$ is a~$\Theta$-patch, then for each $p\in O$ and $q\in Q$ the differential
$\mathrm{d}_{(q,p)}\Theta\colon T_qQ\times T_{p}O\rightarrow T_{q\cdot p}P$ is surjective.
\item[$2.$]
If $N$ is a~$\varphi$-patch and $s_0\colon N \rightarrow P$ a~smooth section, then $s_0(N)$ is a~$\Theta$-patch.
\end{enumerate}
\end{Lemma}

\begin{proof}
1.~Since $O$ is a~$\Theta$-patch, the claim is clear for $q=e$.
If $q$ is arbitrary, then for each $\vec{m}_q\in T_qQ$ we f\/ind some $\vec{q}\in\mathfrak{q}$ such that
$\vec{m}_q=\mathrm{d} L_q\vec{q}$.
Consequently, for $\vec{w}_{p}\in T_{p}P$ we have
\begin{gather*}
\mathrm{d}_{(q,p)}\Theta\left(\vec{m}_q,\vec{w}_{p}\right)=\mathrm{d}_{(q,p)}\Theta(\mathrm{d}
L_q\vec{q},\vec{w}_{p})=\mathrm{d}_{p}L_q\left(\mathrm{d}_{(e,p)}\Theta(\vec{q},\vec{w}_{p})\right).
\end{gather*}
So, since left translation w.r.t.~$\Theta$ is a~dif\/feomorphism, $\mathrm{d}_{p}L_q$ is surjective.

2.~$O:=s_0(N)$ is a~submanifold of $P$ because $s_0$ is an injective immersion. Thus,
by Remark~\ref{bem:Psliceeigensch}.2 it suf\/f\/ices to show that
\begin{gather*}
\dim\big[T_{s_0(x)}O + \mathrm{d}_e\Phi_{s_0(x)}(\mathfrak{g})+ Tv_{s_0(x)}P\big]\geq
\dim[T_{s_0(x)}P]
\qquad\forall\, x\in N.
\end{gather*}
For this, let $x\in N$ and $V'\subseteq \mathfrak{g}$ be a~linear subspace with $V'\oplus \mathfrak{g}_x$ and $T_xM=T_xN \oplus
\mathrm{d}_e\varphi_x(V')$.
Then, we have $T_{s_0(x)}O \oplus \mathrm{d}_e\Phi_{s_0(x)}(V')\oplus Tv_{s_0(x)}P$ because if $\mathrm{d}_xs_0(\vec{v}_x)
+\mathrm{d}_e\Phi_{s_0(x)}(\vec{g}\hspace{1pt}{}')+ \vec{v}_v=0$ for $\vec{v}_x\in T_xN$, $\vec{g}\hspace{1pt}{}'\in V'$ and
$\vec{v}_v\in Tv_{s_0(x)}P$,
\begin{gather*}
0=\mathrm{d}_{s_0(x)}\pi \big(\mathrm{d}_xs_0(\vec{v}_x) +\textstyle\mathrm{d}_e\Phi_{s_0(x)}(\vec{g}\hspace{1pt}{}')+
\vec{v}_v\big)=\vec{v}_x \oplus \mathrm{d}_e\varphi_x(\vec{g}\hspace{1pt}{}')
\end{gather*}
shows $\vec{v}_x \!=\!0$ and $\mathrm{d}_e\phi_{x}(\vec{g}')\!=\!0$, hence $\vec{g}'\!=\!0$ by the choice of $V'$, i.e., $\vec{v}_v\!=\!0$ by assumption.
In particular, $\mathrm{d}_e\phi_{x}(\vec{g}')\!=\!0$ if $\mathrm{d}_e\Phi_{s_0(x)}(\vec{g}')\!=\!0$, hence $\dim[\mathrm{d}_e\Phi_{s_0(x)}(V')]$ $\geq\dim[\mathrm{d}_e\varphi_x(V')]$, from which we obtain
\begin{gather*}
\dim\big[T_{s_0(x)}O + \mathrm{d}_e\Phi_{s_0(x)}(\mathfrak{g})+ Tv_{s_0(x)}P\big]
\geq \dim\big[T_{s_0(x)}O \oplus \mathrm{d}_e\Phi_{s_0(x)}(V')\oplus Tv_{s_0(x)}P\big]
\\
\qquad{}
=\dim[T_xN] + \dim[\mathrm{d}_e\Phi_{s_0(x)}(V')] + \dim[S]
\geq\dim[T_xN] + \dim[\mathrm{d}_e\varphi_x(V')] + \dim[S]\\
\qquad{}
=\dim[P].\tag*{\qed}
\end{gather*}
\renewcommand{\qed}{}
\end{proof}

\section{Characterization of invariant connections}\label{sec:mainth}

In this section, we will use $\Phi$-coverings $\{P_\alpha\}_{\alpha\in I}$ of the bundle $P$ in order to characterize the set of
$\Phi$-invariant connections by families $\{\psi_\alpha\}_{\alpha\in I}$ of smooth maps $\psi_\alpha\colon
\mathfrak{g}\times TP_\alpha \rightarrow \mathfrak{s}$ whose restrictions $\psi_\alpha|_{\mathfrak{g}\times
T_{p_\alpha}P_\alpha }$ are linear and that fulf\/il two additional compatibility conditions.
Here, we will follow the lines of Wang's original approach, which basically means that we generalize the proofs from~\cite{Wang} to the
non-transitive case.
We will proceed in two steps, the f\/irst one being performed in Subsection~\ref{subsec:RedInvConn}.
There, we show that a~$\Phi$-invariant connection gives rise to a~consistent family $\{\psi_\alpha\}_{\alpha\in I}$ of
smooth maps as described above.
We also discuss the situation in~\cite{HarSni} in order to make the two conditions more intuitive.
Then, in Subsection~\ref{subsec:Reconstruc}, we will verify that such families $\{\psi_\alpha\}_{\alpha\in I}$ glue together
to a~$\Phi$-invariant connection on $P$.

\subsection{Reduction of invariant connections}\label{subsec:RedInvConn}

In the following, let $\{P_\alpha\}_{\alpha\in I}$ be a~f\/ixed $\Phi$-covering of $P$ and $\omega$ a~$\Phi$-invariant
connection on $P$.
We def\/ine
\begin{gather*}
\omega_\alpha:=(\Theta^{*}\omega)|_{TQ\times TP_\alpha}\qquad\text{as well as}\qquad\psi_\alpha:=\omega_\alpha|_{\mathfrak{g}\times
TP_\alpha},
\end{gather*}
and for $q'\in Q$ we let $\Co{q'}\colon Q\times P\rightarrow Q\times P$, $(q,p)\mapsto \left(\Co{q'}(q),p\right)$. Finally, we define
\begin{gather*}
\Add{q}(\vec{g}):=\Add{g}(\vec{g})\qquad\forall\, q=(g,s)\in Q,\qquad \forall\, \vec{g}\in \mathfrak{g}.
\end{gather*}
\begin{Lemma}
\label{lemma:omegaalpha}
Let $q\in Q$, $p_\alpha\in P_\alpha$, $p_\beta\in P_\beta$ with\footnote{Recall that, by Convention~\ref{conv:Submnfds}, this actually means $\tau_{P_\beta}(p_\beta)=q\cdot
\tau_{P_\alpha}(p_\alpha)$.} $p_\beta=q\cdot p_\alpha$ and $\vec{w}_{p_\alpha}\in
T_{p_\alpha}P_\alpha$.
Then
\begin{enumerate}\itemsep=0pt
\item[$1)$]
$\omega_\beta(\vec{\eta}\hspace{1pt})=\rho(q)\circ \omega_\alpha(\vec{0}_\mathfrak{q},\vec{w}_{p_\alpha})$ for all $\vec{\eta}\in
TQ\times TP_\beta$ with $\mathrm{d}\Theta(\vec{\eta}\hspace{1pt})=\mathrm{d} L_q\vec{w}_{p_\alpha}$,
\item[$2)$]
$\left(\Co{q}^{*}\omega_\beta\right)\big(\vec{m},\vec{0}_{p_\beta}\big)=\rho(q)\circ
\omega_\alpha\big(\vec{m},\vec{0}_{p_\alpha}\big)$ for all $\vec{m}\in TQ$.
\end{enumerate}
\end{Lemma}

\begin{proof}
1.~Let $\vec{\eta}\in T_{q'}Q\times T_pP_\beta$ for $q'\in Q$.
Then, since\footnote{See end of Subsection~\ref{subsec:InvConn}.} $L_q^{*}\omega = \rho(q)\circ \omega$ for each $q\in Q$
and $q'\cdot p= q\cdot p_\alpha =p_\beta$, we have
\begin{gather*}
\begin{split}
&\omega_\beta(\vec{\eta}\hspace{1pt})=\omega_{q'\cdot p}(\mathrm{d}_{(q', p)}\Theta (\vec{\eta}\hspace{1pt}))=\omega_{p_\beta}(\mathrm{d}
L_q\vec{w}_{p_\alpha}) =(L_q^{*}\omega)_{p_\alpha}(\vec{w}_{p_\alpha})
\\
&\phantom{\omega_\beta(\vec{\eta})}=\rho(q)\circ \omega_{p_\alpha}(\vec{w}_{p_\alpha})=\rho(q)\circ \omega_{p_\alpha}\big(\mathrm{d}_{(e,p_\alpha)}\Theta
\big(\vec{0}_\mathfrak{q},\vec{w}_{p_\alpha}\big)\big)
=\rho(q)\circ
\omega_{\alpha}\big(\vec{0}_\mathfrak{q},\vec{w}_{p_\alpha}\big).
\end{split}
\end{gather*}

2.~For $\vec{m}_{q'} \in T_{q'}Q$ let $\gamma \colon (-\epsilon,\epsilon)\rightarrow Q$ be smooth with
$\dot\gamma(0)=\vec{m}_{q'}$.
Then
\begin{gather*}
\big(\Co{q}^{*}\omega_\beta
\big)_{(q',p_\beta)}\big(\vec{m}_{q'},\vec{0}_{p_\beta}\big)
={\omega_\beta}_{(\Co{q}(q'),p_\beta)}\big(\Add{q}(\vec{m}_{q'}),\vec{0}_{p_\beta}\big)
=\omega_{qq'q^{-1}q\cdot p_\alpha }\left(\dttB{t}{0}q\gamma(t) q^{-1}q\cdot p_\alpha\right)
\\
\hphantom{\big(\Co{q}^{*}\omega_\beta
\big)_{(q',p_\beta)}\big(\vec{m}_{q'},\vec{0}_{p_\beta}\big)}{}
=\left(L_q^{*}\omega\right)_{q'\cdot p_\alpha }\left(\dttB{t}{0}\gamma(t)\cdot p_\alpha\right)
=\rho(q)\circ \omega_{q'\cdot p_\alpha}\left(\mathrm{d}_{(q',p_\alpha)}\Theta\left( \vec{m}_{q'}\right)\right)
\\
\hphantom{\big(\Co{q}^{*}\omega_\beta
\big)_{(q',p_\beta)}\big(\vec{m}_{q'},\vec{0}_{p_\beta}\big)}{}
=\rho(q)\circ {\omega_\alpha}_{(q',p_\alpha)} \big(\vec{m}_{q'},\vec{0}_{p_\alpha}\big).\tag*{\qed}
\end{gather*}
\renewcommand{\qed}{}
\end{proof}

\begin{Corollary}
\label{cor:psialpha}
Let $q\in Q$, $p_\alpha\in P_\alpha$, $p_\beta\in P_\beta$ with $p_\beta=q\cdot p_\alpha$ and $\vec{w}_{p_\alpha}\in
T_{p_\alpha}P_\alpha$.
Then, for $\vec{w}_{p_\beta}\in T_{p_\beta}P_\beta$, $\vec{g}\in \mathfrak{g}$ and $\vec{s}\in \mathfrak{s}$ we have
\begin{enumerate}\itemsep=0pt
\item[$i)$] $\wt{g}(p_\beta) + \vec{w}_{p_\beta}-\wt{s}(p_\beta)=\mathrm{d} L_q\vec{w}_{p_\alpha}
\quad
\Longrightarrow
\quad
\psi_\beta(\vec{g},\vec{w}_{p_\beta})-\vec{s}
=\rho(q)\circ\psi_\alpha\big(\vec{0}_{\mathfrak{g}},\vec{w}_{p_\alpha}\big)$,
\item[$ii)$] $\psi_\beta\big(\Add{q}(\vec{g}),\vec{0}_{p_\beta}\big)=\rho(q)\circ
\psi_\alpha\big(\vec{g},\vec{0}_{p_\alpha}\big)$.
\end{enumerate}
\end{Corollary}

\begin{proof}
$i)$ In general, for $\vec{w}_p\in T_pP$, $\vec{g}\in\mathfrak{g}$ and $\vec{s}\in\mathfrak{s}$ we have
\begin{gather}
\label{eq:ThetaPhi}
\mathrm{d}_{(e,p)}\Theta((\vec{g},\vec{s}\hspace{1pt}),\vec{w}_p)
=\mathrm{d}_{(e,p)}\Phi(\vec{g},\vec{w}_p)-\wt{s}(p)=\wt{g}(p)+\vec{w}_p-\wt{s}(p)
\end{gather}
and, since $\omega$ is a~connection, for $((\vec{g},\vec{s}),\vec{w}_{p_\alpha})\in \mathfrak{q}\times TP_\alpha$ we
obtain{\samepage
\begin{gather}
\begin{split}
& \omega_\alpha((\vec{g},\vec{s}\hspace{1pt}),\vec{w}_{p_\alpha})
=\omega\big(\mathrm{d}_{(e,p_\alpha)}\Phi(\vec{g},\vec{w}_{p_\alpha})-\wt{s}(p_\alpha)\big)
=\omega\big(\mathrm{d}_{(e,p_\alpha)}\Phi(\vec{g},\vec{w}_{p_\alpha})\big)-\vec{s}
\\
&\phantom{\omega_\alpha((\vec{g},\vec{s}),\vec{w}_{p_\alpha})}
=\omega_\alpha\left(\vec{g},\vec{w}_{p_\alpha}\right)-\vec{s}
=\psi_\alpha\left(\vec{g},\vec{w}_{p_\alpha}\right)-\vec{s}.
\end{split} \label{eq:gugu}
\end{gather}
Now,} assume that $\mathrm{d}_e\Phi_{p_\beta}(\vec{g}\hspace{1pt})+\vec{w}_{p_\beta}-\wt{s}(p)=\mathrm{d}
L_q\vec{w}_{p_\alpha}$.
Then $\mathrm{d}_{(e,p_\beta)}\Theta((\vec{g},\vec{s}\hspace{1pt}),\vec{w}_{p_\beta})=\mathrm{d} L_q \vec{w}_{p_\alpha}$
by~\eqref{eq:ThetaPhi} so that $\omega_\beta((\vec{g},\vec{s}),\vec{w}_{p_\beta})=\rho(q)\circ
\omega_\alpha\big(\vec{0}_{\mathfrak{g}},\vec{w}_{p_\alpha}\big)$ by Lemma~\ref{lemma:omegaalpha}.1.
Consequently,
\begin{gather*}
\psi_\beta\left(\vec{g},\vec{w}_{p_\beta}\right)-\vec{s}
\stackrel{\eqref{eq:gugu}}{=}\omega_\beta((\vec{g},\vec{s}\hspace{1pt}),\vec{w}_{p_\beta})
=\rho(q)\circ
\omega_\alpha\big(\vec{0}_\mathfrak{q},\vec{w}_{p_\alpha}\big)\stackrel{\eqref{eq:gugu}}{=}\rho(q)\circ
\psi_\alpha\big(\vec{0}_{\mathfrak{g}},\vec{w}_{p_\alpha}\big).
\end{gather*}

$ii)$ Lemma~\ref{lemma:omegaalpha}.2 yields
\begin{gather*}
\psi_\beta\big(\Add{q}(\vec{g}\hspace{1pt}),\vec{0}_{p_\beta}\big)
=(\Co{q}^{*}\omega_\beta)_{(e,p_\beta)}\big(\vec{g},\vec{0}_{p_\beta}\big)
=\rho(q)\circ (\omega_\alpha)_{(e,p_\alpha)}\big(\vec{g},\vec{0}_{p_\alpha}\big) =\rho(q)\circ
\psi_\alpha\big(\vec{g},\vec{0}_{p_\alpha}\big).\!\!\!\!\tag*{\qed}
\end{gather*}
\renewcommand{\qed}{}
\end{proof}

\begin{Definition}[reduced connection]
A family $\{\psi_\alpha\}_{\alpha\in I}$ of smooth maps $\psi_\alpha\colon \mathfrak{g}\times TP_\alpha \rightarrow
\mathfrak{s}$ which are linear in the sense that $\psi_\alpha|_{\mathfrak{g}\times T_{p_\alpha}P_\alpha }$ is linear for
all $p_\alpha\in P_\alpha$ is called reduced connection w.r.t.~$\{P_\alpha\}_{\alpha\in I}$ if\/f it fulf\/ils the
conditions $i)$ and $ii)$ from Corollary~\ref{cor:psialpha}.
\end{Definition}
\begin{Remark}\label{rem:Psialphaconderkl}\quad
\begin{enumerate}\itemsep=0pt
\item[1)] In particular, Corollary~\ref{cor:psialpha}.$i)$ encodes the following condition
\begin{enumerate}\itemsep=0pt
\item[$a)$] For all $\beta \in I$, $(\vec{g},\vec{s}\hspace{1pt})\in \mathfrak{q}$ and $\vec{w}_{p_\beta}\in
T_{p_\beta}P_\beta$ we have
\begin{gather*}
\wt{g}(p_\beta) +\vec{w}_{p_\beta}-\wt{s}(p_\beta)=0
\quad
\Longrightarrow
\quad
\psi_\beta(\vec{g},\vec{w}_{p_\beta})-\vec{s}=0.
\end{gather*}
\end{enumerate}
\item[2)] Assume that $a)$ is true and let $q\in Q$, $p_\alpha\in P_\alpha$, $p_\beta \in
P_\beta$ with $p_\beta=q\cdot p_\alpha$.
Moreover, assume that we f\/ind elements $\vec{w}_{p_\alpha}\in T_{p_\alpha}P_\alpha$ and
$((\vec{g},\vec{s}\hspace{1pt}),\vec{w}_{p_\beta})\in \mathfrak{q}\times T_{p_\beta}P_\beta$ such that
\begin{gather*}
\mathrm{d}_{(e,p_\beta)}\Theta((\vec{g},\vec{s}\hspace{1pt}),\vec{w}_{p_\beta})=\mathrm{d} L_q\vec{w}_{p_\alpha}
\qquad
\text{and}
\qquad
\psi_\beta(\vec{g},\vec{w}_{p_\beta})-\vec{s}=\rho(q)\circ\psi_\alpha(\vec{0}_{\mathfrak{g}},\vec{w}_{p_\alpha})
\end{gather*}
holds.
Then
$\psi_\beta\big(\vec{g}\hspace{1pt}{}',\vec{w}'_{p_\beta}\big)-\vec{s}\hspace{1pt}{}'
=\rho(q)\circ\psi_\alpha\big(\vec{0}_{\mathfrak{g}},\vec{w}_{p_\alpha}\big)$
holds for each element\footnote{Observe that due to surjectivity of $\mathrm{d}_{(e,p_\beta)}\Phi$ such elements always exist.}
$\big((\vec{g}\hspace{1pt}{}',\vec{s}\hspace{1pt}{}'),\vec{w}'_{p_\beta}\big)\in \mathfrak{q}\times T_{p_\beta}P_\beta$
with\footnote{Recall equation~\eqref{eq:ThetaPhi}.}
$\mathrm{d}_{(e,p_\beta)}\Theta\big((\vec{g}\hspace{1pt}{}',\vec{s}\hspace{1pt}{}'),\vec{w}'_{p_\beta}\big)=\mathrm{d}
L_q\vec{w}_{p_\alpha}$.
In fact, we have
\begin{gather*}
\mathrm{d}_{(e,p_\beta)}\Theta\big((\vec{g}-\vec{g}\hspace{1pt}{}',\vec{s}-\vec{s}\hspace{1pt}{}'),\vec{w}_{p_\beta}-\vec{w}'_{p_\beta}\big)=0,
\end{gather*}
so that by~\eqref{eq:ThetaPhi} condition $a)$ gives
\begin{gather*}
0\stackrel{a)}{=}\psi_\beta(\vec{g}-\vec{g}\hspace{1pt}{}',\vec{w}_{p_\beta}-\vec{w}'_{p_\beta})-(\vec{s}-\vec{s}\hspace{1pt}{}'))
=\big[\psi_\beta(\vec{g},\vec{w}_{p_\beta})-\vec{s}\big]-\big[\psi_\beta(\vec{g}\hspace{1pt}{}',\vec{w}'_{p_\beta})-\vec{s}\hspace{1pt}{}'\big]
\\
\phantom{0}=
\rho(q)\circ\psi_\alpha\big(\vec{0}_{\mathfrak{g}},\vec{w}_{p_\alpha}\big)
-\big[\psi_\beta(\vec{g}\hspace{1pt}{}',\vec{w}'_{p_\beta})-\vec{s}\hspace{1pt}{}'\big].
\end{gather*}
\item[3)] Assume that $\mathrm{d} L_q \vec{w}_{p_\alpha}\in T_{p_\beta}P_\beta$ holds for all $q\in Q$,
$p_\alpha\in P_\alpha$, $p_\beta \in P_\beta$ with $p_\beta=q\cdot p_\alpha$ and all $\vec{w}_{p_\alpha}\in
T_{p_\alpha}P_\alpha$.
Then $\mathrm{d}_{(e,p_\beta)}\Theta\left({\rm d}L_q \vec{w}_{p_\alpha}\right)=\mathrm{d} L_q \vec{w}_{p_\alpha}$ so that it
follows from 2) that in this case we can substitute $i)$ by $a)$ and condition
\begin{enumerate}\itemsep=0pt
\item[$b)$] Let $q\in Q$, $p_\alpha\in P_\alpha$, $p_\beta \in P_\beta$ with $p_\beta=q\cdot
p_\alpha$.
Then
\begin{gather*}
\psi_\beta\big(\vec{0}_{\mathfrak{g}},\mathrm{d}
L_q\vec{w}_{p_\alpha}\big)=\rho(q)\circ\psi_\alpha\big(\vec{0}_{\mathfrak{g}},\vec{w}_{p_\alpha}\big)
\qquad\forall\, \vec{w}_{p_\alpha}\in T_{p_\alpha}P_\alpha.
\end{gather*}
\end{enumerate}
Now, $b)$ looks similar to $ii)$ and makes it plausible that the conditions
$i)$ and $ii)$ from Corollary~\ref{cor:psialpha} together encode the $\rho$-invariance of the corresponding
connection $\omega$.
However, usually there is no reason for $\mathrm{d} L_q \vec{w}_{p_\alpha}$ to be an element of $T_{p_\beta}P_\beta$.
Even for $p_\alpha=p_\beta$ and $q\in Q_{p_\alpha}$ this is usually not true.
Thus, typically there is no way to split up $i)$ into parts whose meaning is more intuitive.
\end{enumerate}
\end{Remark}

Remark~\ref{rem:Psialphaconderkl} immediately proves
\begin{scase}[gauge f\/ixing]
\label{scase:OneSlice}
Let $P_0$ be a~$\Theta$-patch of the bundle $P$ such that $\pi(P_0)$ intersects each $\varphi$-orbit in a~unique point,
and that $\mathrm{d} L_q(T_pP_0)\subseteq T_pP_0$ holds for all $p\in P_0$ and all $q \in Q_p$.
Then, a~corresponding reduced connection consists of one single smooth map $\psi\colon \mathfrak{g}\times TP_0\rightarrow
\mathfrak{s}$, and we have $p=q\cdot p'$ for $q\in Q$, $p,p'\in P_0$ iff $p=p'$ and $q\in Q_p$ holds.
Thus, by Remark~{\rm \ref{rem:Psialphaconderkl}} the two conditions from Corollary~{\rm \ref{cor:psialpha}} are equivalent to:

Let $p\in P_0$, $q=(h,\phi_p(h))\in Q_p$, $\vec{w}_p\in T_pP_0$ and $\vec{g}\in
\mathfrak{g}$, $\vec{s}\in \mathfrak{s}$.
Then
\begin{enumerate}\itemsep=0pt
\item[$i')$] $\wt{g}(p) +\vec{w}_{p}-\wt{s}(p)=0
\quad
\Longrightarrow
\quad
\psi(\vec{g},\vec{w}_{p})-\vec{s}=0$,
\item[$ii')$] $\psi\big(\vec{0}_{\mathfrak{g}},\mathrm{d}
L_q\vec{w}_{p}\big)=\rho(q)\circ\psi\big(\vec{0}_{\mathfrak{g}},\vec{w}_{p}\big)$,
\item[$iii')$]
$\psi\big(\Add{h}(\vec{g}\hspace{1pt}),\vec{0}_{p}\big)=\Add{\phi_p(h)}\circ\psi\big(\vec{g},\vec{0}_{p}\big)$.
\end{enumerate}
\end{scase}

 The next example is a~slight generalization of Theorem~2 in~\cite{HarSni}.
There, the authors assume that $\varphi$ admits only one orbit type so that $\dim[G_x]=l$ holds for all $x\in M$.
Then, they restrict to the situation where one f\/inds a~triple $(U_0,\tau_0,s_0)$ consisting of an open subset $U_0\subseteq
\mathbb{R}^k$ for $k=\dim[M]- [\dim[G]- l]$, an embedding $\tau_0\colon U_0 \rightarrow M$,  and
a~smooth map $s_0\colon U_0\rightarrow P$ with $\pi\circ s_0=\tau_0$ and the addition property that $Q_p$ is the same
for all $p\in \operatorname{\mathrm{im}}[s_0]$.
More precisely, they assume that $G_x$ and the structure group of the bundle are compact.
Then they show the non-trivial fact that $s_0$ can be modif\/ied in such a~way that in addition $Q_p$ is the same for all
$p\in \operatorname{\mathrm{im}}[s_0]$.

Observe that the authors forgot to require that $\operatorname{\mathrm{im}}[\mathrm{d}_x\tau_0] +
\operatorname{\mathrm{im}}\big[\mathrm{d}_e\varphi_{\tau_0(x)}\big]=T_{\tau_0(x)}M$ holds for all $x\in U_0$, i.e., that
$\tau_0(U_0)$ is a~$\varphi$-patch (so that $s_0(U_0)$ is a~$\Theta$-patch).
Indeed, Example~\ref{ex:transinv}.2 shows that this additional condition is crucial.
The next example is a~slight modif\/ication of the result~\cite{HarSni} in the sense that we do not assume $G_x$ and the
structure group to be compact but make the ad hoc requirement that $Q_p$ is the same for all $p\in P_0$.

\begin{Example}[Harnad, Shnider, Vinet]\label{example:SCHSV}
Let $P_0$ be a~$\Theta$-patch of the bundle $P$ such that $\pi(P_0)$ intersects each $\varphi$-orbit in a~unique point. Moreover, assume that the $\Theta$-stabilizer $L:=Q_{p}$ is the same for all $p\in P_0$.
Then, it is clear from~\eqref{eq:staoQ} that $H:=G_{\pi(p)}$ and $\phi:=\phi_{p}\colon H \rightarrow S$ are independent
of the choice of $p\in P_0$.
Finally, we require that
\begin{gather}
\label{eq:dimP}
\dim[P_0]=\dim[M]-[\dim[G]-\dim[H]] \equiv\dim[P]-[\dim[Q]-\dim[H]]
\end{gather}
holds. Now, let $p\in P_0$ and $q=(h,\phi(h))\in Q_p$.
Then, for $\vec{w}_{p}\in T_{p}P_0$ we have
\begin{gather*}
\mathrm{d} L_q\vec{w}_{p}=\dttB{t}{0} \Phi(h,\gamma(t))\cdot \phi^{-1}_{p}(h)
=\dttB{t}{0} [\gamma(t)\cdot\phi_{\gamma(t)}(h)]\cdot \phi^{-1}_{p}(h)
\\
\phantom{\mathrm{d} L_q\vec{w}_{p}}
=\dttB{t}{0} [\gamma(t)\cdot \phi_{p}(h)]\cdot \phi^{-1}_{p}(h)=\vec{w}_{p}
\end{gather*}
for $\gamma\colon (-\epsilon,\epsilon)\rightarrow P_0$ some smooth curve with $\dot\gamma(0)=\vec{w}_{p}$.
Consequently, $\mathrm{d} L_q(T_pP_0)\subseteq T_pP_0$ so that we are in the situation of Case~\ref{scase:OneSlice}.
Here, $ii')$ now reads
$\psi\big(\vec{0}_{\mathfrak{g}},\vec{w}_{p}\big)=\Add{\phi(h)}\circ\psi\big(\vec{0}_{\mathfrak{g}},\vec{w}_{p}\big)$
for all $h\in H$ and $iii')$ does not change.
For $i')$, observe that the Lie algebra $\mathfrak{l}$ of $L$ is contained in the kernel of
$\mathrm{d}_{(e,p_0)}\Theta$; denoting the differential of the restriction of $\Theta$ to $Q\times P_0$ for the moment.
Then, $\mathrm{d}_{(e,p_0)}\Theta$ is surjective by Lemma~\ref{lemma:suralpha}.1 since $P_0$ is a~$\Theta$-patch, so that
\begin{gather*}
\dim\big[\ker\big[\mathrm{d}_{(e,p_0)}\Theta\big]\big]= \dim[Q] +
\dim[P_0]-\dim[P]\stackrel{\eqref{eq:dimP}}{=}\dim[H],
\end{gather*}
hence $\ker[\mathrm{d}_{(e,p)}\Theta]=\mathfrak{l}$ holds for all $p\in P_0$.
Altogether it follows that a~reduced connection w.r.t.~$P_0$ is a~smooth, linear\footnote{In the sense that
$\psi|_{\mathfrak{g}\times T_pP_0}$ is linear for all $p\in P_0$.} map $\psi\colon \mathfrak{g}\times TP_0\rightarrow
\mathfrak{s}$ which fulf\/ils the following three conditions:
\begin{itemize}\itemsep=0pt
\item[$i'')$]
$\psi\big(\vec{h},\vec{0}_{p}\big) \stackrel{\eqref{eq:ThetaPhi}}{=} \mathrm{d}_e\phi\big(\vec{h}\hspace{1pt}\big)
\hspace*{30.3mm}
\forall\, \vec{h}\in \mathfrak{h}, \hspace*{8.9mm}\forall\, p\in P_0$,
\item[$ii'')$]
$\psi\big(\vec{0}_{\mathfrak{g}},\vec{w}\big)=\Add{\phi(h)}\circ
\psi\big(\vec{0}_{\mathfrak{g}},\vec{w}\big)
\hspace*{16mm}
\forall\, h\in H, \qquad\forall\, \vec{w}\in TP_0$,
\item[$iii'')$]
$\psi\big(\Add{h}(\vec{g}),\vec{0}_{p}\big)=\Add{\phi(h)}\circ \psi\big(\vec{g},\vec{0}_{p}\big)
\qquad
\forall\, h\in H, \hspace*{7.7mm}\forall\, \vec{g}\in\mathfrak{g}, \qquad\forall\, p\in P_0$.
\end{itemize}
Then, $\mu:= \psi|_{TP_0}$ and $A_{p_0}(\vec{g}\hspace{1pt}):=\psi\big(\vec{g},\vec{0}_{p_0}\big)$
are the maps that are used for the characterization in Theorem~2 in~\cite{HarSni}.
\end{Example}

\subsection{Reconstruction of invariant connections}\label{subsec:Reconstruc}

Let $\{P_\alpha\}_{\alpha\in I}$ be some fixed~$\Phi$-covering of $P$.
We are going to show that each respective reduced connection $\{\psi_\alpha\}_{\alpha\in I}$ gives rise to a~unique
$\Phi$-invariant connection on $P$.
To this end, for each $\alpha\in I$ we def\/ine the maps $\lambda_\alpha\colon \mathfrak{q}\times TP_\alpha\rightarrow
\mathfrak{s}, ((\vec{g},\vec{s}\hspace{1pt}),\vec{w})\mapsto\psi_\alpha(\vec{g},\vec{w})-\vec{s}$ and
\begin{gather*}
\begin{split}
&\omega_\alpha\colon\quad TQ\times TP_\alpha  \rightarrow  \mathfrak{s},
\\
&\phantom{\omega_\alpha\colon\quad {}}
\big(\vec{m}_q,\vec{w}_{p_\alpha}\big) \mapsto  \rho(q)\circ \lambda_\alpha\left(\mathrm{d}
L_{q^{-1}}\vec{m}_q,\vec{w}_{p_\alpha}\right)
\end{split}
\end{gather*}
for $\vec{m}_q\in T_qQ$ and $\vec{w}_{p_\alpha}\in T_{p_\alpha}P_\alpha$.

\begin{Lemma}
\label{lemma:lambda}
Let $q\in Q$, $p_\alpha\in P_\alpha$, $p_\beta\in P_\beta$ with $p_\beta=q\cdot p_\alpha$ and $\vec{w}_{p_\alpha}\in
T_{p_\alpha}P_\alpha$.
Then
\begin{enumerate}\itemsep=0pt
\item[$1)$]
$\lambda_\beta(\vec{\eta}\hspace{1pt})=\rho(q)\circ \lambda_\alpha\big(\vec{0}_{\mathfrak{q}},\vec{w}_{p_\alpha}\big)$ for all
$\vec{\eta}\in \mathfrak{q}\times T_{p_\beta}P$ with $\mathrm{d}\Theta_{(e,p_\beta)}(\vec{\eta}\hspace{1pt})=\mathrm{d}
L_q\vec{w}_{p_\alpha}$,
\item[$2)$]
$\lambda_\beta\big(\Add{q}(\vec{q}\hspace{1pt}),\vec{0}_{p_\beta}\big)=\rho(q)\circ
\lambda_\alpha\big(\vec{q},\vec{0}_{p_\alpha}\big)$ for all $\vec{q}\in \mathfrak{q}$.
\end{enumerate}
For each $\alpha \in I$ we have
\begin{enumerate}\itemsep=0pt
\item[$3)$] $\ker\big[\lambda_\alpha|_{\mathfrak{q}\times T_{p_\alpha}P_\alpha}\big]\subseteq
\ker\big[\mathrm{d}_{(e,p_\alpha)}\Theta\big]$ for all $p_\alpha\in P_\alpha$,
\item[$4)$] the map $\omega_\alpha$ is
the unique $\mathfrak{s}$-valued $1$-form on $Q\times P_\alpha$ which extends $\lambda_\alpha$ and for which we have
$L_q^{*}\omega_\alpha=\rho(q)\circ \omega_\alpha$ for all $q\in Q$.
\end{enumerate}
\end{Lemma}

\begin{proof}
1.~Write $\vec{\eta}=((\vec{g},\vec{s}\hspace{1pt}),\vec{w}_{p_\beta})$ for $\vec{g}\in\mathfrak{g}$, $\vec{s}\in\mathfrak{s}$ and
$\vec{w}_{p_\beta}\in T_{p_\beta}P_\beta$.
Then
\begin{gather*}
\wt{g}(p_\beta)
+\vec{w}_{p_\beta}-\wt{s}(p_\beta)\stackrel{\eqref{eq:ThetaPhi}}{=}\mathrm{d}\Theta_{(e,p_\beta)}(\vec{\eta})=\mathrm{d}
L_q\vec{w}_{p_\alpha}
\end{gather*}
so that from condition $i)$ in Corollary~\ref{cor:psialpha} we obtain
\begin{gather*}
\lambda_\beta(\vec{\eta}\hspace{1pt})=\psi_\beta(\vec{g},\vec{w}_{p_\beta})-\vec{s}
=\rho(q)\circ\psi_\alpha\big(\vec{0}_{\mathfrak{g}},\vec{w}_{p_\alpha}\big)
=\rho(q)\circ\lambda_\alpha\big(\vec{0}_{\mathfrak{q}},
\vec{w}_{p_\alpha}\big).
\end{gather*}

2.~Let $\vec{q}=(\vec{g},\vec{s}\hspace{1pt})$ for $\vec{g}\in\mathfrak{g}$ and $\vec{s}\in \mathfrak{s}$.
Then, by Corollary~\ref{cor:psialpha}.$ii)$ we have
\begin{gather*}
\lambda_\beta\big(\Add{q}(\vec{q}\hspace{1pt}),\vec{0}_{p_\beta}\big)=\psi_\beta\big(\Add{q}(\vec{g}\hspace{1pt}),\vec{0}_{p_\beta}\big)\!-
\Add{q}(\vec{s}\hspace{1pt})
=\rho(q)\circ [\hspace{1pt}\psi_\alpha\big(\vec{g},\vec{0}_{p_\alpha}\big)\!- \vec{s}\hspace{2pt}] =\rho(q) \circ
\lambda_\alpha\big(\vec{q},\vec{0}_{p_\alpha}\big).
\end{gather*}

3.~This follows from the f\/irst part for $\alpha=\beta$, $q=e$ and $\vec{w}_{p_\alpha}=\vec{0}_{p_\alpha}$.

4.~By def\/inition we have $\omega_\alpha|_{\mathfrak{q}\times TP_\alpha}=\lambda_\alpha$, and for the pullback property we
calculate
\begin{gather*}
\left(L_{q'}^{*}\omega_\alpha\right)_{(q,p_\alpha)}\big(\vec{m}_q,\vec{w}_{p_\alpha}\big)
={\omega_\alpha}_{(q'q,p_\alpha)}\big(\mathrm{d}
L_{q'}\vec{m}_q,\vec{w}_{p_\alpha}\big)
=\rho\left(q'q\right)\circ\lambda_\alpha\left(\mathrm{d} L_{q^{-1}q'^{-1}}\mathrm{d}
L_{q'}\vec{m}_q,\vec{w}_{p_\alpha}\right)
\\
\hphantom{\left(L_{q'}^{*}\omega_\alpha\right)_{(q,p_\alpha)}\big(\vec{m}_q,\vec{w}_{p_\alpha}\big)}{}
=\rho\left(q'\right)\circ\rho(q)\circ \lambda_\alpha\left(\mathrm{d}
L_{q^{-1}}\vec{m}_q,\vec{w}_{p_\alpha}\right)
=\rho\left(q'\right)\circ {\omega_\alpha}_{(q,p_\alpha)}(\vec{m}_q,\vec{w}_{p_\alpha}),
\end{gather*}
where $q,q'\in Q$ and $\vec{m}_q\in T_qQ$.
For uniqueness, let $\omega$ be another $\mathfrak{s}$-valued 1-form on $Q\times P_\alpha$ whose restriction to
$\mathfrak{q}\times TP_\alpha$ is $\lambda_\alpha$ and that fulf\/ils $L_q^{*}\omega=\rho(q)\circ \omega$ for all $q\in Q$.
Then
\begin{gather*}
\omega_{(q,p_\alpha)}\left(\vec{m}_q,\vec{w}_{p_\alpha}\right)={\omega}_{(q,p_\alpha)}\left(\mathrm{d} L_q\circ
\mathrm{d} L_{q^{-1}}\vec{m}_q,\vec{w}_{p_\alpha}\right) =(L_q^{*}\omega)_{(e,p_\alpha)}\left(\mathrm{d}
L_{q^{-1}}\vec{m}_q,\vec{w}_{p_\alpha}\right)
\\
\hphantom{\omega_{(q,p_\alpha)}\left(\vec{m}_q,\vec{w}_{p_\alpha}\right)}{}
=\rho(q)\circ \omega_{(e,p_\alpha)}(\mathrm{d} L_{q^{-1}}\vec{m}_q,\vec{w}_{p_\alpha}) = \rho(q)\circ
\lambda_\alpha\left(\mathrm{d} L_{q^{-1}}\vec{m}_q,\vec{w}_{p_\alpha}\right)
\\
\hphantom{\omega_{(q,p_\alpha)}\left(\vec{m}_q,\vec{w}_{p_\alpha}\right)}{}
=\omega_\alpha(\mathrm{d} L_{q^{-1}}\vec{m}_q,\vec{w}_{p_\alpha}).
\end{gather*}
Finally, smoothness of $\omega_\alpha$ is an easy consequence of smoothness of the maps $\rho$, $\lambda_\alpha$ and
$\mu \colon TQ\rightarrow \mathfrak{q}$, $\vec{m}_q \mapsto \mathrm{d} L_{q^{-1}}\vec{m}_q$ with $\vec{m}_q\in T_qQ$.
For this, observe that $\mu={\rm d}\tau \circ \kappa$ for $\tau\colon Q\times Q\rightarrow Q$, $(q,q')\mapsto q^{-1}q'$ and
$\kappa\colon TQ\rightarrow TQ\times TQ$, $\vec{m}_q\mapsto \big(\vec{0}_q,\vec{m}_q\big)$ for $\vec{m}_q\in T_qQ$.
\end{proof}

So far, we have shown that each reduced connection $\{\psi_\alpha\}_{\alpha \in I}$ gives rise to uniquely
determined maps $\{\lambda_\alpha\}_{\alpha \in I}$ and $\{\omega_\alpha\}_{\alpha \in I}$.
In the f\/inal step, we will construct a~unique $\Phi$-invariant connection $\omega$ from the data
$\{(P_\alpha,\lambda_\alpha)\}_{\alpha\in I}$.
Here, uniqueness and smoothness of $\omega$ will follow from uniqueness and smoothness of the maps $\omega_\alpha$.

\begin{Proposition}\label{prop:reconstr}
There is one and only one $\mathfrak{s}$-valued $1$-form $\omega$ on $P$ with
$\omega_\alpha=(\Theta^{*}\omega)|_{TQ\times TP_\alpha}$ for all $\alpha \in I$. This $1$-form is a $\Phi$-invariant connection on~$P$.
\end{Proposition}

\begin{proof}
For uniqueness, we have to show that the values of such an $\omega$ are uniquely determined by the maps $\omega_\alpha$.
To this end, let $p\in P$, $\alpha\in I$ and $p_\alpha\in P_\alpha$ be such that $p=q\cdot p_\alpha$ holds for some $q\in Q$.
By Lemma~\ref{lemma:suralpha}.1 for $\vec{w}_p\in T_pP$ we f\/ind some $\vec{\eta}\in T_qQ\times T_{p_\alpha}P_\alpha$
with $\vec{w}_p=\mathrm{d}_{(q,p_\alpha)}\Theta(\vec{\eta})$, so that uniqueness follows from
\begin{gather*}
\omega_p(\vec{w}_p)=\omega_{q\cdot p_\alpha}\left(\mathrm{d}_{(q,p_\alpha)}\Theta(\vec{\eta}\hspace{1pt})\right)
=(\Theta^{*}\omega)_{(q,p_\alpha)}(\vec{\eta}\hspace{1pt}) =\omega_\alpha(\vec{\eta}\hspace{1pt}).
\end{gather*}
For existence, let $\alpha\in I$ and $p_\alpha\in P_\alpha$.
Due to surjectivity of $\mathrm{d}_{(e,p_\alpha)}\Theta$ and Lemma~\ref{lemma:lambda}.3, there is a~(unique) map
$\widehat{\lambda}_{p_\alpha}\colon T_{p_\alpha}P\rightarrow \mathfrak{s}$ with
\begin{gather}
\label{eq:lambda}
\widehat{\lambda}_{p_\alpha}\circ \mathrm{d}_{(e,p_\alpha)}\Theta=\lambda_\alpha\big|_{\mathfrak{q}\times
T_{p_\alpha}P_\alpha}.
\end{gather}
Let $\widehat{\lambda}_\alpha \colon \bigsqcup_{p_\alpha\in P_\alpha} T_{p_\alpha} P\rightarrow \mathfrak{s}$ denote the
(unique) map whose restriction to $T_{p_\alpha}P$ is $\widehat{\lambda}_{p_\alpha}$ for each $p_\alpha\in P_\alpha$.
Then $\lambda_\alpha=\widehat{\lambda}_\alpha\circ \mathrm{d} \Theta|_{\mathfrak{q}\times TP_\alpha}$ and we construct
the connection $\omega$ as follows.
For $p\in P$ we choose some $\alpha\in I$ and $(q,p_\alpha)\in Q\times P_\alpha$ such that $q\cdot p_\alpha=p$ and
def\/ine
\begin{gather}
\label{eq:defomega}
\omega_p\big(\vec{w}_p\big):=\rho(q)\circ \widehat{\lambda}_\alpha\left(\mathrm{d} L_{q^{-1}}\big(\vec{w}_p\big)\right)
\qquad
\forall\, \vec{w}_p\in T_pP.
\end{gather}
We have to show that this depends neither on $\alpha\in I$ nor on the choice of $(q,p_\alpha)\in Q\times P_\alpha$.
For this, let $p_\alpha\in P_\alpha$, $p_\beta \in P_\beta$ and $q\in Q$ with $p_\beta=q\cdot p_\alpha$.
Then for $\vec{w}\in T_{p_\alpha}P$ we have $\vec{w}=\mathrm{d}\Theta(\vec{q},\vec{w}_{p_\alpha})$ for some
$(\vec{q},\vec{w}_{p_\alpha})\in \mathfrak{q}\times T_{p_\alpha}P_\alpha$, and since $\mathrm{d}
L_q\vec{w}_{p_\alpha}\in T_{p_\beta}P$, there is $\vec{\eta}\in \mathfrak{q}\times T_{p_\beta}P_\beta$ such that
$\mathrm{d}_{(e,p_\beta)}\Theta(\vec{\eta}\hspace{1pt})=\mathrm{d} L_q\vec{w}_{p_\alpha}$ holds.
It follows from the conditions~1 and~2 in Lemma~\ref{lemma:lambda} that
\begin{gather}
\begin{split}
& \widehat{\lambda}_\beta(\mathrm{d} L_q\vec{w})
=\widehat{\lambda}_\beta((\mathrm{d} L_q\circ\mathrm{d}\Theta)(\vec{q},\vec{w}_{p_\alpha}))
=\widehat{\lambda}_\beta\big((\mathrm{d} L_q\circ\mathrm{d}\Theta)
\big(\vec{q},\vec{0}_{p_\alpha}\big)\big)+\widehat{\lambda}_\beta\big(\mathrm{d}L_q\vec{w}_{p_\alpha}\big)
\\
& \phantom{\widehat{\lambda}_\beta(\mathrm{d} L_q\vec{w})}\hspace{-1.5mm}
\stackrel{\eqref{eq:thirdstep}}{=}\widehat{\lambda}_\beta\circ \mathrm{d}\Theta \big(
\Add{q}(\vec{q}\hspace{1pt}),\vec{0}_{p_\beta}\big)+\widehat{\lambda}_\beta\circ \mathrm{d}\Theta(\vec{\eta}\hspace{1pt})
\\
& \phantom{\widehat{\lambda}_\beta(\mathrm{d} L_q\vec{w})}\hspace{-1.5mm}
\stackrel{\eqref{eq:lambda}}{=}\lambda_\beta\big(\Add{q}(\vec{q}\hspace{1pt}),\vec{0}_{p_\beta}\big)+\lambda_\beta(\vec{\eta}\hspace{1pt})
=\rho(q)\circ \lambda_\alpha\big(\vec{q},\vec{0}_{p_\alpha}\big)+\rho(q)\circ
\lambda_\alpha\big(\vec{0}_{\mathfrak{q}},\vec{w}_{p_\alpha}\big)
\\
& \phantom{\widehat{\lambda}_\beta(\mathrm{d} L_q\vec{w})}
=\rho(q)\circ\lambda_\alpha(\vec{q},\vec{w}_{p_\alpha}) =\rho(q)\circ \widehat{\lambda}_\alpha\circ
\mathrm{d}\Theta(\vec{q},\vec{w}_{p_\alpha}) =\rho(q)\circ \widehat{\lambda}_\alpha(\vec{w}),
\end{split}\label{eq:wohldefs}
\end{gather}
where for the third equality we have used that
\begin{gather}
\begin{split}
 \left(\mathrm{d} L_q\circ\mathrm{d}\Theta\right)\big(\vec{q},\vec{0}_{p_\alpha}\big)
&=\dttB{t}{0}\hspace{1pt}q\cdot(\exp(t\vec{q}\hspace{1pt})\cdot p_\alpha)
\\
&=\dttB{t}{0}\hspace{1pt}\Co{q}(\exp(t\vec{q}\hspace{1pt}))\cdot p_\beta=\mathrm{d}\Theta\big(\Add{q}(\vec{q}\hspace{1pt}),\vec{0}_{p_\beta}\big).
\end{split}
\label{eq:thirdstep}
\end{gather}
Consequently, if $\wt{q}\cdot p_\beta=p$ with $(\wt{q},p_\beta)\in Q\times P_\beta$ for some $\beta \in I$, then
$p_\beta=(q^{-1}\wt{q})^{-1}\cdot p_\alpha$ and well-def\/inedness follows from
\begin{gather*}
\rho(\wt{q})\circ \widehat{\lambda}_\beta\left(\mathrm{d}
L_{{\wt{q}}^{-1}}(\vec{w}_p)\right)=\rho(q)\circ\rho\big(q^{-1}\wt{q}\hspace{1pt}\big)\circ\widehat{\lambda}_\beta\left(\mathrm{d}
L_{(q^{-1}\wt{q})^{-1}}\big( \mathrm{d} L_{q^{-1}}\vec{w}_p\big)\right)
\\
\phantom{\rho(\wt{q})\circ \widehat{\lambda}_\beta\left(\mathrm{d}
L_{{\wt{q}}^{-1}}(\vec{w}_p)\right)}
{} =\rho(q)\circ \widehat{\lambda}_\alpha\big(\mathrm{d} L_{q^{-1}}\vec{w}_p\big),
\end{gather*}
where the last step is due to~\eqref{eq:wohldefs} with $\vec{w}=\mathrm{d} L_{q^{-1}}\vec{w}_p\in T_{p_\alpha}P$.
Next, we show that $\omega$ fulf\/ils the pullback property.
For this, let $(\vec{m},\vec{w}_{p_\alpha})\in T_qQ\times T_{p_\alpha}P_\alpha$.
Then
\begin{gather*}
(\Theta^{*}\omega)\left(\vec{m}_q,\vec{w}_{p_\alpha}\right)
=\omega_{q\cdot p_\alpha}\left(\mathrm{d}\Theta(\vec{m}_q,\vec{w}_{p_\alpha})\right) \stackrel{\eqref{eq:defomega}}{=}\rho(q)\circ
\widehat{\lambda}_\alpha\left(\mathrm{d} L_{q^{-1}}\mathrm{d}\Theta(\vec{m}_q,\vec{w}_{p_\alpha})\right)
\\
\phantom{(\Theta^{*}\omega)\left(\vec{m}_q,\vec{w}_{p_\alpha}\right)}
=\rho(q)\circ \widehat{\lambda}_\alpha\circ \mathrm{d}\Theta\left(\mathrm{d} L_{q^{-1}}
\vec{m}_q,\vec{w}_{p_\alpha}\right) \stackrel{\eqref{eq:lambda}}{=}\rho(q)\circ \lambda_\alpha\left(\mathrm{d}
L_{q^{-1}} \vec{m}_q,\vec{w}_{p_\alpha}\right)
\\
\phantom{(\Theta^{*}\omega)\left(\vec{m}_q,\vec{w}_{p_\alpha}\right)}
=\omega_\alpha(\vec{m}_q,\vec{w}_{p_\alpha}).
\end{gather*}
In the third step, we have used that $L_{q^{-1}}\circ \Theta= \Theta(L_{q^{-1}}(\cdot),\cdot)$.
Finally, we have to verify that $\omega$ is a~$\Phi$-invariant smooth connection.
For this, let $p\in P$ and $(\wt{q},p_\alpha)\in Q\times P_\alpha$ with $p=\wt{q}\cdot p_\alpha$.
Then, for $q\in Q$ and $\vec{w}_p\in T_pP$ we have
\begin{gather*}
\begin{split}
&\left(L_q^{*}\omega\right)_{p}(\vec{w}_{p})=\omega_{q\cdot p}\left(\mathrm{d} L_q \vec{w}_{p}\right)
=\omega_{(q\wt{q})\cdot p_\alpha}\left(\mathrm{d} L_q \vec{w}_{p}\right)
\\
&\phantom{\left(L_q^{*}\omega\right)_{p}(\vec{w}_{p})}
=\rho(q)\circ\rho\left(\wt{q}\right)\circ \widehat{\lambda}_\alpha\left(\mathrm{d}
L_{\wt{q}^{-1}}\vec{w}_{p}\right) =\rho(q)\circ \omega_{p}(\vec{w}_{p}),
\end{split}
\end{gather*}
hence
\begin{gather*}
R_s^{*}\omega\phantom{}=L_{\left(e,s^{-1}\right)}^{*}\omega=\rho\big(\big(e,s^{-1}\big)\big)\circ \omega
=\Add{s^{-1}}\circ \omega,
\\
L_g^{*}\omega =L_{(g,e)}^{*}\omega=\rho((g,e))\circ \omega=\omega.
\end{gather*}
Thus, it remains to show smoothness of $\omega$, and that $\omega_p(\widetilde{s}(p))=\vec{s}$ holds for all $p\in P$ and
all $\vec{s}\in \mathfrak{s}$.
For the second property, let $p=q\cdot p_\alpha$ for $(q,p_\alpha)\in Q\times P_\alpha$.
Then $q=(g,s)$ for some $g\in G$ and $s\in S$ and we obtain
\begin{gather*}
\omega_{p}(\widetilde{s}(p))=\rho(q)\circ \widehat{\lambda}_\alpha\left(\mathrm{d} L_{q^{-1}}\widetilde{s}(q\cdot
p_\alpha)\right)
=\rho(q)\circ \widehat{\lambda}_\alpha\left(\dttB{t}{0}\hspace{2pt} p_\alpha\cdot
\left(\Co{s^{-1}}(\exp(t\vec{s}\hspace{1pt})\right)\right)
\\
\phantom{\omega_{p}(\widetilde{s}(p))}
=\rho(q)\circ\widehat{\lambda}_\alpha
\big(\mathrm{d}\Theta\big(\Add{s^{-1}}(\vec{s}\hspace{1pt}),\vec{0}_{p_\alpha}\big)\big)\!
=\Add{s}\!\circ \lambda_\alpha\big(\Add{s^{-1}}(\vec{s}\hspace{1pt}),\vec{0}_{p_\alpha}\!\big)\!
=\Add{s}\circ\Add{s^{-1}}(\vec{s}\hspace{1pt})=\vec{s}.
\end{gather*}
For smoothness, let $p_\alpha\in P_\alpha$ and choose a~submanifold $Q'$ of $Q$ through $e$, an open neighbourhood
$P'_\alpha\subseteq P_\alpha$ of $p_\alpha$, and an open subset $U\subseteq P$ such that the restriction
$\widehat{\Theta}:=\Theta|_{Q' \times P'_\alpha}$ is a~dif\/feomorphism to $U$.
Then $p_\alpha \in U$ because $e\in Q'$, hence
\begin{gather*}
\omega|_U=\widehat{\Theta}^{-1*} \big[\widehat{\Theta}^{*}\omega\big]
=\widehat{\Theta}^{-1*} \big[(\Theta^{*}\omega)|_{TQ\times
TP_\alpha}\big] =\widehat{\Theta}^{-1*} \omega_\alpha.
\end{gather*}
Since $\omega_\alpha$ is smooth and $\widehat{\Theta}$ is a~dif\/feomorphism, $\omega|_U$ is smooth as well.
Finally, if $p=q\cdot p_\alpha$ holds for $q\in Q$, then $L_q(U)$ is an open neighbourhood of $p$ and
\begin{gather*}
\omega|_{L_q(U)}=\big(L_{q^{-1}}^{*}\left(L_q^{*}\omega\right)\big)\big|_{L_q(U)}= \rho(q)\circ
\big(L_{q^{-1}}^{*}\omega\big)\big|_{L_q(U)} =\rho(q)\circ L_{q^{-1}}^{*} \left(\omega|_U\right)
\end{gather*}
is smooth because $\omega|_U$ and $L_{q^{-1}}$ are smooth.
\end{proof}

 Corollary~\ref{cor:psialpha} and Proposition~\ref{prop:reconstr} now prove
\begin{Theorem}\label{th:InvConnes}
Let~$G$ be a~Lie group of automorphisms of the principal fibre bundle $P$.
Then, for each $\Phi$-covering $\{P_\alpha\}_{\alpha\in I}$ of $P$
the assignment
\begin{gather*}
	\omega \mapsto \{(\Phi^*\omega)|_{\mathfrak{g}\times TP_\alpha}\}_{\alpha\in I}
\end{gather*}
is a~bijection between
the $\Phi$-invariant connections on $P$ and the
reduced connections that correspond to $\{P_\alpha\}_{\alpha\in I}$.
\end{Theorem}

 As already mentioned in the remarks following Case~\ref{scase:OneSlice}, the second part of the next
example shows the importance of the transversality condition
\begin{gather*}
\operatorname{\mathrm{im}}[\mathrm{d}_x\tau_0] +
\operatorname{\mathrm{im}}\big[\mathrm{d}_e\varphi_{\tau_0(x)}\big]=T_{\tau_0(x)}M
\qquad\forall\, x\in U_0
\end{gather*}
for the formulation in~\cite{HarSni}.
\begin{Example}[(semi-)homogeneous connections]\label{ex:transinv}\quad
\begin{enumerate}\itemsep=0pt
\item
Let $P=X \times S$ for an $n$-dimensional $\mathbb{R}$-vector space $X$ and an arbitrary structure group~$S$.
Moreover, let $G \subseteq X$ be a~linear subspace of dimension $1\leq k\leq n$ acting via
\begin{gather*}
\Phi\colon G\times P
\rightarrow P,\quad (g,(x,\sigma))\mapsto(g+x,\sigma).
\end{gather*}
If $W$ is an algebraic complement of~$G$ in $X$ and $P_0:=W\times \{e_S\}\subseteq P$, then $P_0$ is a~$\Phi$-covering because $\Theta\colon (G\times S)\times P_0\rightarrow P$ is a~dif\/feomorphism and each $\varphi$-orbit intersects $W$ in
a~unique point.
Consequently, identifying $G$ with its Lie algebra $\mathfrak{g}$, the $\Phi$-invariant connections on $P$ are in bijection with the smooth maps $\psi\colon G\times
TW\rightarrow \mathfrak{s}$ for which $\psi_w:=\psi|_{G \times T_wW}$ is linear for all $w\in W$.
This is because the conditions $i)$ and $ii)$ from Corollary~\ref{cor:psialpha} give no further
restrictions in this case.
It is straightforward to see\footnote{Pull back $\omega^\psi$ by $\Theta$ and restrict it to $\mathfrak{g}\times TP_0$.} that the $\Phi$-invariant connection that corresponds to $\psi$ is given by
\begin{gather}
\label{eq:transinv}
\omega^\psi_{(x,s)}(\vec{v}_x,\vec{\sigma}_s)
=\Add{s^{-1}}\circ\psi_{\mathrm{pr}_W(x)}\big(\mathrm{pr}_G(\vec{v}_x),\mathrm{pr}_W(\vec{v}_x)\big)+\mathrm{d}
L_{s^{-1}}(\vec{\sigma}_s)
\end{gather}
for $(\vec{v}_x,\vec{\sigma}_s)\in T_{(x,s)}P$.
\item
Let $(P,\pi,M,S)$ be a principal fibre bundle with Lie group of automorphisms $(G,\Phi)$. Then, for $(U_0,\tau_0,s_0)$ a triple as in \cite{HarSni}\footnote{See also the discussions following Case \ref{scase:OneSlice}.}, Theorem 2 in~\cite{HarSni} states that each smooth $\ovl{\psi}\colon \mathfrak{g}\times TU_0\rightarrow \mathfrak{s}$ for which $\ovl{\psi}|_{\mathfrak{g}\times T_{x}U_0}$ is linear for all $x\in U_0$, and that fulfils the three conditions from Example \ref{example:SCHSV} can be written as $(\Phi^*\ovl{\omega})|_{\mathfrak{g}\times TU_0}$ for some (even unique)
 invariant connection $\ovl{\omega}$ on $P$.

  We consider the situation of the previous part, whereby we let
\begin{gather*}
X=\mathbb{R}^2\qquad G=\mathrm{span}_\mathbb{R}(\vec{e}_1)\qquad W=\mathrm{span}_\mathbb{R}(\vec{e}_2)\qquad \text{and}\qquad P_0=W\times \{e\}.
\end{gather*}
Now, we are going to construct $(U_0,\tau_0,s_0)$ and $\ovl{\psi}$ in such a way that the above statement is wrong:
\begingroup
\setlength{\leftmarginii}{13pt}
\begin{itemize}\itemsep=0pt
\item
First, we fix $0\neq \vec{s}\in \mathfrak{s}$ and define $\omega$ by \eqref{eq:transinv} for $\psi\colon \mathfrak{g}\times TP_0 \rightarrow \mathfrak{s}$ the map
    \begin{align*}
      \psi_{y}(\lambda \cdot\vec{e}_1,\mu\cdot \vec{e}_2):= \mu \cdot f(y)\cdot\vec{s}\qquad \text{for}\qquad (\lambda \cdot\vec{e}_1,\mu \cdot\vec{e}_2)\in \mathfrak{g}\times T_{(y\cdot\vec{e}_{2},e)}P_0
    \end{align*}    	
    with $f(0):=0$ and $f(y):=1\slash \sqrt[3]{y}$ if $y\neq 0$. Then, $\omega$ is easily seen to be smooth on $P':=Z\times S$ for $Z:= \{(x,y)\in \mathbb{R}^2\:|\: y\neq 0\}$, but it is not smooth at $((x,0),e)$ because
    \begin{align*}
      \omega_{((x,y),e)}\big(\big(\vec{0},\vec{e}_2\big),\vec{0}_{\mathfrak{s}}\big)=\psi_y\big(\vec{0},\vec{e}_2\big)=f(y)\cdot \vec{s}\qquad \forall\: y\in \mathbb{R}.
    \end{align*}
	Even more: there cannot exist any smooth invariant connection $\omega$ on $P$ which coincides on $P'$ with $\omega$, just because $\lim_{y\rightarrow 0}f(y)\cdot \vec{s}$ does not exist.
\item
Now, we let $U_0:=\mathbb{R}$, $\tau_0 \colon U_0 \rightarrow \mathbb{R}^2$, $t\mapsto \big(t,t^3\big)$ and $s_0\colon t \mapsto (\tau_0(t),e)$. Then, $(U_0,\tau_0,s_0)$ fulfils the conditions from \cite{HarSni}, but we have\footnote{
Thus, $(U_0,s_0)$ cannot be a $\Theta$-patch by the second part of Remark \ref{bem:Psliceeigensch}.3.}
\begin{gather*}
\mathrm{im}[\mathrm{d}_0\tau_0] + \mathrm{im}\hspace{-1pt}\big[\mathrm{d}_e\varphi_{\tau_0(0)}\big] = \mathrm{span}_{\mathbb{R}}(\vec{e}_1)\neq T_0X =T_0\mathbb{R}^2=\mathbb{R}^2.
\end{gather*}
      As a consequence, $\ovl{\psi}\colon \mathfrak{g}\times TU_0\rightarrow \mathfrak{s}$ defined by $\ovl{\psi}_t:=(\Phi^*\omega)|_{\mathfrak{g}\times T_t U_0}$ is smooth, because for $t\neq 0$ and $r\in T_t U_0 =\mathbb{R}$ we have
    \begin{align*}
      \ovl{\psi}_{t}(\lambda\hspace{1pt} \vec{e}_1,r)
&= (\Phi^*\omega)_{(e,s_0(t))}(\lambda\vec{e}_1,\mathrm{d}_t s_0(r))
      =\big(\Phi^*\omega\big)\left(\lambda\hspace{1pt} \vec{e}_1,r\cdot \vec{e}_1 + 3t^2 r\cdot \vec{e}_2\right)\\
      &=\omega_{((t,t^3),e)}\big((\lambda+r)\cdot \vec{e}_1 + 3t^2 r\cdot \vec{e}_2,\vec{0}_{\mathfrak{s}}\big)\\
      &=\psi_{t^3}\!\left(\left(\lambda+r\right)\cdot \vec{e}_1,3t^2 r\cdot \vec{e}_2\right)= 3 t r\cdot \vec{s}
    \end{align*}
    as well as $\ovl{\psi}_{0}(\lambda\hspace{1pt} \vec{e}_1,r)=0$ if $t=0$. For the first step, keep in mind that
	\begin{align*}
    (\Phi^*\omega)|_{\mathfrak{g}\times T_tU_0}(\vec{g},r)=(\Phi^*\omega)(\vec{g},\mathrm{d}_t s_0(r))
    \end{align*}
    holds by Convention \ref{conv:Submnfds}.2. Since $\omega$ fulfils the algebraic properties of an invariant connection, $\ovl{\psi}$ fulfils the algebraical properties from Example \ref{example:SCHSV}.
    \item
    It remains to show that there is no smooth invariant connection $\ovl{\omega}$ on $P$ for which $\ovl{\psi}=(\Phi^*\ovl{\omega})|_{\mathfrak{g}\times TU_0}$ holds. This, however, follows from the first point as such an $\ovl{\omega}$ necessarily had to coincide  on $P'$ with $\omega$.

    In fact, let $U'_0:=\mathbb{R}_{\neq 0}$, and $\tau_0'\colon U_0'\rightarrow Z$, $t\mapsto \big(t,t^3\big)$ as well as $s_0'\colon t\mapsto(\tau_0'(t),e)$ be defined as above. Then, $(U_0',s_0')$ is a $\Theta$-patch as we have removed the point $0\in U_0$ for which transversality fails. Thus, $(U_0',s_0')$ is a $\Phi$-covering of $P'$, so that
    \begin{gather*}
    	(\Phi^*\ovl{\omega})|_{\mathfrak{g}\times T U'_0}=\ovl{\psi}|_{\mathfrak{g}\times T U'_0}=(\Phi^*\omega)|_{\mathfrak{g}\times T U'_0}
    \end{gather*}
    implies $\ovl{\omega}=\omega$ on $P'$.
\end{itemize}
\endgroup

\end{enumerate}
\end{Example}

\section{Particular cases and applications}\label{sec:PartCases}

In the f\/irst part of this section, we will consider $\Phi$-coverings of $P$ arising from the induced action~$\varphi$ on
the base manifold~$M$ of~$P$.
Then, we discuss the case where $\Phi$ acts via gauge transformations on~$P$, which will lead us to a~straightforward generalization of the description of connections by consistent families of local 1-forms
on~$M$.
In the second part, we discuss the (almost) f\/ibre transitive case, and deduce Wang's original theorem~\cite{Wang} from
Theorem~\ref{th:InvConnes}.
Finally, we will consider the situation where $P$ is trivial, and give examples in loop quantum gravity.

\subsection[$\Phi$-coverings and the induced action]{$\boldsymbol{\Phi}$-coverings and the induced action}%\label{subsec:GCovIndAct}

Let $(G,\Phi)$ be a~Lie group of automorphisms of the principal f\/ibre bundle $P$.
According to Lemma~\ref{lemma:mindimslice}, for each $x\in M$ there is a~$\varphi$-patch (with minimal dimension) $M_x$
with $x\in M$.
Consequently, we find an open neighbourhood $M'_x\subseteq M_x$ of $x$ and a~local section $s_x\colon U\rightarrow P$
with $M'_x\subseteq U$ for $U$ an open neighbourhood of $M$.
Let $I\subseteq M$ be a~subset such that\footnote{It is always possible to choose $I=M$.} each $\varphi$-orbit
intersects at least one of the sets $M_x$ for some $x\in I$.
Then, it is immediate from Lemma~\ref{lemma:suralpha}.2 that $\{s_x(M'_x)\}_{x\in I}$ is a~$\Phi$-covering of $P$.
More generally, we have

\begin{Corollary}
\label{cor:reductions}
Let $(P,\pi,M,S)$ be a~principal fibre bundle and $(G,\Phi)$ a~Lie group of automorphisms of $P$.
Denote by $(M_\alpha,s_\alpha)_{\alpha\in I}$ a~family consisting of a~collection of $\varphi$-patches
$\{M_\alpha\}_{\alpha\in I}$ and smooth sections\footnote{This is that $\pi\circ s_\alpha= \mathrm{id}_{M_\alpha}$.}
$s_\alpha\colon M_\alpha\rightarrow P$.
Then, the sets $P_\alpha:=s_\alpha(M_\alpha)$ are $\Theta$-patches.
They provide a~$\Phi$-covering of $P$ iff each $\varphi$-orbit intersects at least one patch $M_\alpha$.
\end{Corollary}

\begin{proof}
This is immediate from Lemma~\ref{lemma:suralpha}.2.
\end{proof}

 We now consider the case where $(G,\Phi)$ is a~Lie group of gauge transformations of $P$, i.e.,
$\varphi_g=\mathrm{id}_M$ for all $g\in G$.
Here, we show that Theorem~\ref{th:InvConnes} can be seen as a~generalization of the description of smooth connections by means of
consistent families of local 1-forms on the base manifold~$M$.
For this, let $\{U_\alpha\}_{\alpha\in I}$ be an open covering of~$M$ and $\{s_\alpha\}_{\alpha\in I}$ a~family of smooth
sections \mbox{$s_\alpha\colon U_\alpha \rightarrow P$}.
We def\/ine the open sets $U_{\alpha\beta}:=U_\alpha \cap U_\beta$ and consider the smooth maps $\delta_{\alpha\beta}\colon G\times
U_{\alpha\beta}\rightarrow S$ determined by $s_\beta(x)=\Phi(g,s_\alpha(x))\cdot
\delta_{\alpha\beta}(g,x)$, and for which $\delta_{\alpha\beta}(g,x)=\phi^{-1}_{s_\alpha(x)}(g)
\cdot\delta_{\alpha\beta}(e,x)$ holds. Finally, we let $\mu_{\alpha\beta}(g,\vec{v}_x):= \mathrm{d}
L_{\delta^{-1}_{\alpha\beta}(g,x)}\circ \mathrm{d}_x\delta_{\alpha\beta}(g,\cdot)(\vec{v}_x)$ for $\vec{v}_x\in
T_xU_{\alpha\beta}$ and $g\in G$.
Then, we have
\begin{scase}[Lie groups of gauge transformations]
\label{scase:GaugeTransf}
Let $(G,\Phi)$ be a~Lie group of gauge transformations of the principal fibre bundle $(P,\pi,M,S)$.
Then, the $\Phi$-invariant connections on $P$ are in bijection with the families $\{\chi_\alpha\}_{\alpha\in I}$ of
$\mathfrak{s}$-valued $1$-forms $\chi_\alpha\colon U_\alpha\rightarrow\mathfrak{s}$ for which we have
\begin{gather}
\label{eq:consistgauge}
\chi_\beta(\vec{v}_x)=\big(\Add{\delta_{\alpha\beta}(g,x)}\circ\chi_\alpha\big)(\vec{v}_x)
+\mu_{\alpha\beta}(g,\vec{v}_x)
\qquad\forall\, \vec{v}_x\in T_xU_{\alpha\beta},\qquad\forall\, g\in G.
\end{gather}
\begin{proof}
By Corollary~\ref{cor:reductions} $\{s_\alpha(U_\alpha)\}_{\alpha\in I }$ is a~$\Phi$-covering of $P$.
So, let $\{\psi_\alpha\}_{\alpha\in I}$ be a~reduced connection w.r.t.\ this covering.
We f\/irst show that condition $i)$ from Corollary~\ref{cor:psialpha} implies
\begin{gather*}%\label{eq:eq1}
\psi_\beta\big(\vec{g},\vec{0}_{p}\big) =\mathrm{d}_e\phi_p(\vec{g}\hspace{1pt})
\qquad
\forall\,
\vec{g}\in \mathfrak{g}, \qquad\forall\, p\in s_\beta(U).
\end{gather*}
For this observe that condition $a)$ from Remark~\ref{rem:Psialphaconderkl} means that for all $\beta\in I$, $p
\in s_\beta(U_\beta)$, $\vec{w}_p\in T_ps_\beta(U_\beta)$ and $\vec{g}\in \mathfrak{g}$, $\vec{s}\in \mathfrak{s}$ we
have
\begin{gather*}
\mathrm{d}_e\Phi_p(\vec{g}\hspace{1pt}) + \vec{w}_p -\widetilde{s}(p)=0
\quad
\Longrightarrow
\quad
\psi_\beta(\vec{g},\vec{w}_p)-\vec{s}=0.
\end{gather*}
Now, $T_ps_\beta(U_\beta)$ is complementary to $Tv_pP$ and $\operatorname{\mathrm{im}}[\mathrm{d}_e\Phi_p]\subseteq
\ker[\mathrm{d}_p\pi]$ so that $a)$ is the same as
\begin{enumerate}\itemsep=0pt
\item[$a')$] $\mathrm{d}_e\Phi_p(\vec{g}\hspace{1pt})=\widetilde{s}(p)
\quad
\Longrightarrow
\quad
\psi_\beta\big(\vec{g},\vec{0}_p\big)=\vec{s}$
\qquad
for  \ $\vec{g}\in \mathfrak{g}$, $\vec{s}\in \mathfrak{s}$ and all $p\in P_\beta$.
\end{enumerate}
But, since $G_x=G$ for all $x\in M$, this just means\footnote{$\mathrm{d}_e\Phi_p(\vec{g}\hspace{1pt})-
\widetilde{s}(p)=0$ if\/f \:$(\vec{g},\vec{s}\hspace{1pt})\in \mathfrak{q}_p$ if\/f
 $\vec{s}=\mathrm{d}_e\phi_p(\vec{g})$.} $\psi_\beta\big(\vec{g},\vec{0}_{p}\big)
=\mathrm{d}_e\phi_p(\vec{g})$ for all $\vec{g}\in \mathfrak{g}$ and already implies Condition~$ii)$
from Corollary~\ref{cor:psialpha} as $\phi_p$ is a~Lie group homomorphism.
Consequently, we can ignore this condition in the following.
Now, we have $p_\beta = q\cdot p_\alpha$ for $q\in Q$, $p_\alpha \in P_\alpha$, $p_\beta \in P_\beta$ if\/f
$\pi(p_\alpha)=\pi(p_\beta)=x\in U_{\alpha\beta}$ and $q=\big(g,\delta^{-1}_{\alpha\beta}(g,x)\big)$.
Consequently, the left hand side of condition $i)$ from Corollary~\ref{cor:psialpha} reads
\begin{gather*}
\wt{g}(s_\beta(x))+\mathrm{d}_x s_\beta (\vec{v}_\beta) - \wt{s}(s_\beta(x))= \big(\mathrm{d} L_g\circ \mathrm{d}
R_{\delta_{\alpha\beta}(g,x)}\circ \mathrm{d}_x s_\alpha\big) (\vec{v}_\alpha),
\end{gather*}
where $\vec{v}_\alpha, \vec{v}_\beta \in T_xM$ and $g\in G$.
This is true for $\vec{v}_\alpha=\vec{v}_\beta=\vec{v}_x$, $\vec{g}=0$ and $\vec{s}=\mu_{\alpha\beta}(g,\vec{v}_x)$,
which follows from
\begin{gather*}
\mathrm{d}_x s_\beta (\vec{v}_\beta)
=\mathrm{d}_x \big[L_g\circ R_{\delta_{\alpha\beta}(g,\cdot)}\circ s_\alpha\big](\vec{v}_x)
\\
\phantom{\mathrm{d}_x s_\beta (\vec{v}_\beta)}
=\mathrm{d} L_g\big[\mathrm{d}_{s_\alpha(x)}R \big(\mathrm{d}_x\delta_{\alpha\beta}(g,\cdot)(\vec{v}_x)\big)
+\mathrm{d} R_{\delta_{\alpha\beta}(g,x)}(\mathrm{d}_x s_\alpha(\vec{v}_x))\big],
\\
\wt{s}(s_\beta(x))=\dttB{t}{0}L_g \circ R_{\delta_{\alpha\beta}(g,x)\cdot \exp(t\vec{s}\hspace{1pt})}(s_\alpha(x))
\\
\phantom{\wt{s}(s_\beta(x))}
=\mathrm{d} L_g\big[\mathrm{d}_{s_\alpha(x)}R\big(\mathrm{d}L_{\delta_{\alpha\beta}(g,x)}(\vec{s}\hspace{1pt})\big)\big]
=\mathrm{d} L_g\big[\mathrm{d}_{s_\alpha(x)}R\big(\mathrm{d}_x\delta_{\alpha\beta}(g,\cdot)(\vec{v}_x)\big)\big].
\end{gather*}
Consequently, by Corollary~\ref{cor:psialpha}.$i)$ and for
\begin{gather*}
(\psi_\alpha \circ \mathrm{d}_x
s_{\alpha})(\vec{v}_x):= \psi_\alpha\big(\vec{0}_{\mathfrak{g}},\mathrm{d}_x s_\alpha(\vec{v}_x)\big)\qquad\forall\: \vec{v}_x\in
T_xU_{\alpha\beta}
\end{gather*}
we have
\begin{gather}
\label{eq:gaugetraf}
\psi_\beta\big(\vec{0}_{\mathfrak{g}},\mathrm{d}_x s_\beta(\vec{v}_x)\big)=\big(\Add{\delta_{\alpha\beta}(g,x)} \circ
\psi_\alpha \circ \mathrm{d}_x s_{\alpha}\big)(\vec{v}_x)+\mu_{\alpha\beta}(g,\vec{v}_x)
\end{gather}
for all $g\in G$ and all $\vec{v}_x\in T_xU_{\alpha\beta}$.
Due to part 2)
in Remark~\ref{rem:Psialphaconderkl} the condition $i)$ from Corollary~\ref{cor:psialpha}
now gives no further restrictions, so that for $\chi_\beta:=\psi_\beta \circ \mathrm{d} s_\beta$ we have
\begin{gather*}
\psi_\beta(\vec{g}, \mathrm{d}_xs_\beta(\vec{v}_x))=\mathrm{d}_e\phi_{s_\beta(x)}(\vec{g}\hspace{1pt})+ \chi_\beta(\vec{v}_x)
\qquad\forall\, \vec{g}\in \mathfrak{g},
\qquad\forall\, \vec{v}_x\in T_xM,
\qquad\forall\, x\in U_\beta.
\end{gather*}
Then, $\psi_\beta$ is uniquely determined by $\chi_\beta$ for each $\beta\in I$, so that~\eqref{eq:gaugetraf} yields the
consistency condition~\eqref{eq:consistgauge} for the maps $\{\chi_\alpha\}_{\alpha\in I}$.
\end{proof}
\end{scase}
\begin{Example}[trivial action] If~$G$ acts trivially, then for each $x\in U_{\alpha\beta}$ we have
\begin{gather*}
\delta_{\alpha\beta}(g,x)=\phi_{s_\alpha(x)}^{-1}(g)\cdot \delta_{\alpha\beta}(e,x)=\delta_{\alpha\beta}(e,x).
\end{gather*}
Thus, $\delta_{\alpha\beta}$ is independent of $g\in G$, so that Case~\ref{scase:GaugeTransf} just reproduces the description of smooth connections by means of consistent families
of local 1-forms on the base manifold~$M$.
\end{Example}

\subsection{(Almost) fibre transitivity}%\label{subsec:AlmFTC}

In this subsection we discuss the situation where~$M$ admits an element that is contained in the closure of each
$\varphi$-orbit.
For instance, this holds for all $x\in M$ if each $\varphi$-orbit is dense in~$M$ and, in particular, is true for f\/ibre
transitive actions.

\begin{scase}[almost f\/ibre transitivity]
\label{scase:slicegleichredcluster}
Let $x\in M$ be contained in the closure of each $\varphi$-orbit and let $p\in F_x$.
Then, each $\Theta$-patch $P_0\subseteq P$ with $p\in P_0$ is a~$\Phi$-covering of $P$.
Hence, the $\Phi$-invariant connections on $P$ are in bijection with the smooth maps $\psi\colon \mathfrak{g}\times
TP_0\rightarrow \mathfrak{s}$ for which $\psi|_{\mathfrak{g}\times T_pP_0}$ is linear for all $p\in P_0$ and that fulfil
the two conditions from Corollary~{\rm \ref{cor:psialpha}}.
\end{scase}

\begin{proof}\looseness=1
It suf\/f\/ices to show that $\pi\left(P_0\right)$ intersects each $\varphi$-orbit $[o]$.
Since $P_0$ is a~$\Theta$-patch, there is an open neighbourhood $P'\subseteq P_0$ of $p$ and a~submanifold~$ Q'$ of~$Q$
through $(e_G,e_S)$ such that $\Theta|_{Q' \times P'}$ is a~dif\/feomorphism to an open subset~$U\subseteq P$.
Then $\pi(U)$ is an open neighbourhood of~$\pi(p)$ and by assumption we have $[o]\cap \pi(U)\neq \varnothing$ for each
$[o]\in M\slash G$.
Consequently, for $[o]\in M\slash G$ we f\/ind $\wt{p}\in U$ with $\pi(\wt{p})\in [o]$.
Let $\wt{p}=\Theta((g',s'),p')$ for $((g',s'),p')\in Q'\times P'$.
Then
\begin{gather*}
[o]\ni\pi(\wt{p})=\pi\left(\Phi(g',p')\cdot s'\right)=\varphi(g',\pi(p'))\in [\pi(p')]
\end{gather*}
shows that $[o]=[\pi(p')]$ holds, hence $\pi\left(P_0\right)\cap [o]\neq \varnothing$.
\end{proof}

 The next example to Case~\ref{scase:slicegleichredcluster} shows that evaluating the conditions $i)$
and $ii)$ from Corollary~\ref{cor:psialpha} at one single point can be suf\/f\/icient to verify non-existence of
invariant connections.

\begin{Example}[general linear group]\label{ex:Bruhat} \quad
\begin{enumerate}\itemsep=0pt
\item
Let $P:={\rm GL}(n,\mathbb{R})$ and $G=S=B \subseteq {\rm GL}(n,\mathbb{R})$ the subgroup of upper triangular mat\-rices.
Moreover, let $S_n\subseteq {\rm GL}(n,\mathbb{R})$ be the group of permutation matrices.
Then, $P$ is a~principal f\/ibre bundle with base manifold $M:=P\slash S$, structure group $S$ and projection map
\mbox{$\pi\colon P\rightarrow M$}, $p\mapsto [p]$.
Moreover,~$G$ acts via automorphisms on $P$ by $\Phi(g,p):=g\cdot p$, and we have the Bruhat decomposition
\begin{gather*}
{\rm GL}(n,\mathbb{R})=\bigsqcup_{w\in S_n}B w B.
\end{gather*}
Then, $M=\bigsqcup_{w\in S_n}G\cdot\pi(w)$, $G\cdot \pi(e)=\pi(e)$ and $\pi(e)\in \ovl{G\cdot\pi(w)}$ for all $w\in S_n$.
Now, $\operatorname{\mathrm{im}}[\mathrm{d}_e\Theta_e]=\mathfrak{g}$ since $\mathrm{d}_e\Theta_e(\vec{g}\hspace{1pt})=\vec{g}$ for
all $\vec{g}\in \mathfrak{g}$.
Moreover, $\mathfrak{g}={\mathrm{span}_\mathbb{R}}\{E_{ij} \:|\: 1\leq i\leq j\leq n\}$, so that
$V:={\mathrm{span}_\mathbb{R}}\{E_{ij} \:|\: 1\leq j<i\leq n\}$ is an algebraic complement of $\mathfrak{g}$ in
$T_eP=\mathfrak{gl}(n,\mathbb{R})$.
By Lemma~\ref{lemma:mindimslice}.2 we f\/ind a~patch $H\subseteq P$ through $e$ with $T_eH=V$, and due to
Case~\ref{scase:slicegleichredcluster} this is a~$\Phi$-covering.
\item
A closer look at the point $e\in P$ shows that there cannot exist any $\Phi$-invariant connection on
${\rm GL}(n,\mathbb{R})$.
In fact, if $\psi\colon \mathfrak{g}\times TH \rightarrow \mathfrak{s}$ is a~reduced connection w.r.t.~$H$, for
$\vec{w}:=\vec{0}_e$ and $\vec{g}=\vec{s}$ we have
\begin{gather*}
\wt{g}(e)+\vec{w} -\widetilde{s}(e)=\vec{g}+\vec{w}-\vec{s}=0.
\end{gather*}
Thus, condition $i)$ from Corollary~\ref{cor:psialpha} gives $\psi\big(\vec{g},\vec{0}_e\big)-\vec{g}=0$, hence
$\psi\big(\vec{g},\vec{0}_e\big)=\vec{g}$ for all $\vec{g}\in \mathfrak{g}$.
Now, $q\cdot e =e$ if\/f $q=(b,b)$ for some $b\in B$.
Let
\begin{gather*}
V\ni\vec{h}:=E_{n1}, \qquad B\ni b:= e + E_{1n}, \qquad \mathfrak{g}\ni \vec{g}:=E_{11}- E_{1n}-
E_{nn}.
\end{gather*}
Then, $\wt{g}(e)+\vec{h}=\vec{g}+\vec{h}=b\vec{h} b^{-1}=\mathrm{d} L_q\vec{h}$, so that condition $i)$ yields
\begin{gather*}
\psi\big(\vec{g},\vec{h}\hspace{1pt}\big)=\rho(q)\circ \psi\big(\vec{0}_{\mathfrak{g}},\vec{h}\hspace{1pt}\big)=\Add{b}\circ
\psi\big(\vec{0}_{\mathfrak{g}},\vec{h}\hspace{1pt}\big),
\end{gather*}
hence $\vec{g}+\left[\mathrm{id}-\Add{b}\right]\circ \psi\big(\vec{0}_{\mathfrak{g}},\vec{h}\big)=0$.
But, $(\vec{g})_{11}=1$ and
\begin{gather*}
\big(\psi\big(\vec{0}_{\mathfrak{g}},\vec{h}\hspace{1pt}\big)-\Add{b}\circ
\psi
\big(\vec{0}_{\mathfrak{g}},\vec{h}\hspace{1pt}\big)\big)_{11}
=\big(\psi\big(\vec{0}_{\mathfrak{g}},\vec{h}\hspace{1pt}\big)\big)_{11} -\big(\psi\big(\vec{0}_{\mathfrak{g}},\vec{h}\hspace{1pt}\big)\big)_{11} =0,
\end{gather*}
so that $\psi$ cannot exist.
\end{enumerate}
\end{Example}

\begin{Corollary}\label{cor:SliceFibTran}
If $\Phi$ is fibre transitive, then $\{p\}$ is a~$\Phi$-covering for all $p\in P$.
\end{Corollary}

\begin{proof}
It suf\/f\/ices to show that $\{\pi(p)\}$ is a~$\varphi$-patch, since then $\{p\}$ is a~$\Theta$-patch by
Corollary~\ref{cor:reductions}, and a~$\Phi$-covering by Case~\ref{scase:slicegleichredcluster}.
This, however, is clear from Remark~\ref{rem:patch}.1.
In fact, if $x:=\pi(p)$, then by general theory we know that~$M$ is dif\/feomorphic to $G\slash G_{x}$ via $\vartheta
\colon [g]\mapsto \varphi(g,x)$ and that for each $[g]\in G\slash G_{x}$ we f\/ind an open neighbourhood $U\subseteq
G\slash G_{x}$ of $[g]$ and a~smooth section $s\colon U\rightarrow G$.
Then, surjectivity of $\mathrm{d}_e\varphi_x$ is clear from surjectivity of $\mathrm{d}_{[e]}\vartheta$ and
\begin{gather*}
\mathrm{d}_e\varphi_{x}\circ \mathrm{d}_{[e]}s= \mathrm{d}_{[e]}(\varphi_{x} \circ s)=
\mathrm{d}_{[e]}\varphi(s(\cdot),x)=\mathrm{d}_{[e]}\vartheta,
\end{gather*}
showing that $T_xM=\mathrm{d}_e\varphi_x(\mathfrak{g})$ holds.
\end{proof}

Let $\varphi$ be transitive and $p\in P$.
Then, $\{p\}$ is a~$\Phi$-covering by Corollary~\ref{cor:SliceFibTran} and $T_p\{p\}$ is the zero vector space.
Moreover, we have $p_\alpha=q\cdot p_\beta$ if\/f $p_\alpha=p_\beta=p$ and $q\in Q_p$.
It follows that a~reduced connection w.r.t.~$\{p\}$ can be seen as a~linear map $\psi\colon\mathfrak{g}\rightarrow
\mathfrak{s}$ that fulf\/ils the following two conditions:{\samepage
\begin{itemize}\itemsep=0pt
\item
$\mathrm{d}_e\Theta_p(\vec{g},\vec{s}\hspace{1pt})=0
\quad
\Longrightarrow
\quad
\psi(\vec{g}\hspace{1pt})=\vec{s}$
\qquad
for \  $\vec{g}\in \mathfrak{g}$, $\vec{s}\in \mathfrak{s}$,
\item
$\psi\big(\Add{q}(\vec{g}\hspace{1pt})\big)=\rho(q)\circ \psi(\vec{g}\hspace{1pt})
\hspace*{23mm}\forall\, q\in Q_p,
\qquad\forall\, \vec{g}\in \mathfrak{g}$.
\end{itemize}}

Since $\ker[d_e\Theta_p]=\mathfrak{q}_p$, we have shown
\begin{scase}[Hsien-Chung Wang,~\cite{Wang}]
\label{th:wang}
Let $(G,\Phi)$ be a~fibre transitive Lie group of automorphisms of $P$.
Then, for each $p\in P$ there is a~bijection between the $\Phi$-invariant connections on $P$ and the linear maps
$\psi\colon \mathfrak{g}\rightarrow \mathfrak{s}$ that fulfil
\begin{enumerate}\itemsep=0pt
\item[\rm a)]
$\psi\big(\vec{h}\hspace{1pt}\big)=\mathrm{d}_e\phi_p\big(\vec{h}\hspace{1pt}\big)\hspace*{17.4mm}\forall\, \vec{h}\in \mathfrak{g}_{\pi(p)}$,
\item[\rm b)] $\psi\circ\Add{h}=\Add{\phi_p(h)}\circ \psi\qquad\forall\, h\in G_{\pi(p)}$.
\end{enumerate}
This bijection is explicitly given by $\omega\mapsto
\Phi_p^{*}\omega$.
\end{scase}

\begin{Example}\label{ex:eukl}\quad
\begin{enumerate}\itemsep=0pt
\item
\textbf{Homogeneous connections.}
In the situation of Example~\ref{ex:transinv} let $k=n$ and $X=\mathbb{R}^n$.
Then, $\Phi$ is f\/ibre transitive, and for $p=(0,e)$ we have $G_{\pi(p)}=\{e\}$ as well as $\mathfrak{g}_{\pi(p)}=\{0\}$.
Thus, the reduced connections w.r.t.~$\{p\}$ are just the linear maps $\psi\colon \mathbb{R}^n \rightarrow
\mathfrak{s}$, and the corresponding homogeneous connections are given by
\begin{gather*}
{\omega^\psi}_{(x,s)}(\vec{v}_x,\vec{\sigma}_s)=\Add{s^{-1}}\circ\psi(\vec{v}_x)+\mathrm{d} L_{s^{-1}}(\vec{\sigma}_s)
\qquad\forall\, (\vec{v}_x,\vec{\sigma}_s)\in T_{(x,s)}P.
\end{gather*}
\item
\textbf{Homogeneous isotropic connections.}
Let $P=\mathbb{R}^3 \times {\rm SU}(2)$ and $\varrho \colon {\rm SU}(2) \rightarrow {\rm SO}(3)$ be the universal covering map.
We consider the semi direct product $E:=\mathbb{R}^3 \rtimes_\varrho {\rm SU}(2)$ whose multiplication is given by
$(v,\sigma)\cdot_\varrho (v',\sigma'):=(v+\varrho(\sigma)(v'),\sigma\sigma')$ for all $(v,\sigma),(v',\sigma)\in E$.
Since $E$ equals $P$ as a~set, we can def\/ine the action $\Phi$ of $E$ on $P$ just by $\cdot_\varrho$.
Then, $E$ is a~Lie group which resembles the euclidean one, and it follows from Wang's theorem that the $\Phi$-invariant
connections are of the form (see, e.g.,\ Appendix A.3 in~\cite{InvConnLQG}) \begin{gather*}
\omega^c_{(x,s)}(\vec{v}_x,\vec{\sigma}_x)= c\, \Add{s^{-1}}[{\mathfrak{z}}(\vec{v}_x)]+s^{-1}\vec{\sigma}_s
\qquad\forall\, (\vec{v}_x,\vec{\sigma}_s)\in T_{(x,s)}P.
\end{gather*}
Here, $c$ runs over $\mathbb{R}$ and ${\mathfrak{z}}\colon \textstyle\sum\limits_{i=1}^3 v^i \vec{e}_i\rightarrow
\sum\limits_{i=1}^3 v^i \tau_i$ with matrices
\begin{gather*}
\qquad
\tau_1:=
\begin{pmatrix}
0 & -\mathrm{i}
\\
-\mathrm{i} & 0
\end{pmatrix},
\qquad
\tau_2:=
\begin{pmatrix}
0 & -1
\\
1 & 0
\end{pmatrix},
\qquad
\tau_3:=
\begin{pmatrix}
-\mathrm{i} & 0
\\
0 & \mathrm{i}
\end{pmatrix},
\end{gather*}
and $\{\vec{e}_1,\vec{e}_2,\vec{e}_3\}$ the standard basis in $\mathbb{R}^3$.
\end{enumerate}
\end{Example}

We close this section with a~remark concerning the relations between sets of invariant connections that
correspond to dif\/ferent lifts of the same Lie group action on the base manifold of a~principal f\/ibre bundle.
\begin{Remark}
Let $P$ be a~principal f\/ibre bundle and $\Phi,\Phi'\colon G\times P\rightarrow P$ two Lie groups of automorphisms
with $\varphi=\varphi'$.
Then, the respective sets of invariant connections can dif\/fer signif\/icantly.
In fact, in the situation of the second part of Example~\ref{ex:eukl} let $\Phi'((v,\sigma),(x,s)):=(v+
\varrho(\sigma)(x),s)$.
Then, $\varphi'=\varphi$ and Appendix~\ref{subsec:DifferentLifts} shows that
$\omega_0(\vec{v}_x,\vec{\sigma}_s):=s^{-1}\vec{\sigma}_s$ for $(\vec{v}_x,\vec{\sigma}_s)\in T_{(x,s)}P$ is the only
$\Phi'$-invariant connection on $P$.
\end{Remark}

\subsection{Trivial bundles~-- applications to LQG}\label{sec:ApplTrivB}

In this section, we will determine the set of spherically symmetric connections on $\mathbb{R}^3\times {\rm SU}(2)$ to be used for the
description of spherically symmetric gravitational systems (such as black holes) in the framework of loop quantum gravity.
To this end, we reformulate Theorem~\ref{th:InvConnes} for trivial bundles.

The spherically symmetric connections on $P=\mathbb{R}^3\times {\rm SU}(2)$ are such connections, invariant under the action
$\Phi\colon {\rm SU}(2) \times P\rightarrow P$, $(\sigma,(x,s))\mapsto (\sigma(x),\sigma s)$.
Since $\Phi$ is not f\/ibre transitive, we cannot use Case~\ref{th:wang} for the necessary calculations.
Moreover, it is not possible to apply the results from~\cite{HarSni} (see Example~\ref{example:SCHSV}) because the
$\varphi$-stabilizer of $x=0$ equals ${\rm SU}(2)$ whereas that of each $x\in \mathbb{R}^3\backslash \{0\}$ is given by the
maximal torus $T_x:=\{\exp(t {\mathfrak{z}}(x) \:|\: t\in \mathbb{R})\}\subseteq {\rm SU}(2)$.
Of course, we could ignore the origin and consider the bundle $\mathbb{R}^3\backslash \{0\}\times {\rm SU}(2)$ together with
the $\Phi$-covering $\{\lambda \cdot \vec{e}_1\:|\: \lambda \in \mathbb{R}_{>0}\}$.
This, however, is a~dif\/ferent situation because an invariant connection on $\mathbb{R}^3\backslash \{0\}\times {\rm SU}(2)$ is
not necessarily extendible to an invariant connection on $\mathbb{R}^3\times {\rm SU}(2)$ as the next example
illustrates\footnote{See also the remarks following Example~\ref{bsp:Rotats}, as well as the connection $\omega$ constructed in Example~\ref{ex:transinv}.2.}.
\begin{Example}\label{ex:OnePoint}\quad
\begin{enumerate}\itemsep=0pt
\item
Let $S$ be a~Lie group and $P=\mathbb{R}^n\times S$.
We consider the action $\Phi\colon \mathbb{R}_{>0}\times P \rightarrow P$, $(\lambda,(x,s))\mapsto (\lambda x,s)$ and
claim that the only $\Phi$-invariant connection is given by
\begin{gather*}
\omega_0(\vec{v}_x,\vec{\sigma}_s):=\mathrm{d}_sL_{s^{-1}}(\vec{\sigma}_s)
\qquad\forall\, (\vec{v}_x,\vec{\sigma}_s)\in T_{(x,e)}P.
\end{gather*}
In fact, $P_\infty:=\mathbb{R}^n\times \{e\}$ is a~$\Phi$-covering of $P$ by Corollary~\ref{cor:reductions}, and it is
straightforward to see (cf.\ Remark~\ref{rem:Psialphaconderkl}.3) that condition $i)$ from Corollary~\ref{cor:psialpha} is
equivalent to the conditions~$a)$ and~$b)$ from Remark~\ref{rem:Psialphaconderkl}.
Let $\psi\colon \mathfrak{g}\times TP_\infty$ be a~reduced connection w.r.t.~$P_{\infty}$ and def\/ine
$\psi_x:=\psi|_{\mathfrak{g}\times T_{(x,e)}}$.

Since the exponential map $\exp\colon \mathfrak{g}\rightarrow \mathbb{R}_{>0}$ is just given by $\mu \mapsto
\mathrm{e}^{\mu}$ for $\mu \in \mathbb{R}=\mathfrak{g}$, we have $\wt{g}((x,e))=\vec{g}\cdot x\in
T_{(x,e)}P_\infty$ for $\vec{g}\in \mathfrak{g}$.
Then, for $\vec{w}:=-\vec{g}\cdot x \in T_{(x,e)}P_\infty$ from $a)$ we obtain
\begin{gather}
\label{eq:conda}
\psi_x\big(\vec{g},\vec{0}\hspace{1pt}\big)=\psi_x\big(\vec{0}_{\mathfrak{g}},\vec{g}\cdot x\big)
\qquad\forall\, \vec{g}\in \mathfrak{g},\qquad\forall\, x\in \mathbb{R}^n.
\end{gather}
In particular, $\psi_0\big(\vec{g},\vec{0}\hspace{1pt}\big)=0$, and since $Q_{(0,e)}=\mathbb{R}_{>0}\times \{e\}$, for
$q=(\lambda,e)$ condition $b)$ yields
\begin{gather*}
\lambda\,\psi_{0}\big(\vec{0}_{\mathfrak{g}},\vec{w}\big)=\psi_{0}\big(\vec{0}_{\mathfrak{g}},\lambda
\vec{w}\big)\stackrel{b)}{=}\psi_{0}\big(\vec{0}_{\mathfrak{g}},\vec{w}\big)
\qquad\forall\, \lambda> 0,\qquad\forall\, \vec{w} \in T_{(0,e)}P_\infty,
\end{gather*}
hence $\psi_0=0$.
Analogously, for $x\neq 0$, $\vec{w} \in T_{(\lambda x,e)}P_\infty$, $\lambda>0$ and $q=(\lambda,e)$, we obtain
\begin{gather*}
\lambda\,\psi_{\lambda x}\big(\vec{0}_{\mathfrak{g}},\vec{w}\big)=\psi_{\lambda
x}\big(\vec{0}_{\mathfrak{g}},\mathrm{d} L_q(\vec{w})\big)\stackrel{b)}{=}\rho(q)\circ
\psi_{x}\big(\vec{0}_{\mathfrak{g}},\vec{w}\big)= \psi_{x}\big(\vec{0}_{\mathfrak{g}},\vec{w}\big),
\end{gather*}
i.e., $\psi_{\lambda x}\big(\vec{0}_{\mathfrak{g}},\vec{w}\big)=\frac{1}{\lambda}
\psi_{x}\big(\vec{0}_{\mathfrak{g}},\vec{w}\big)$.
Here, in the second step, we have used the canonical identif\/ication of the linear spaces $T_{(x,e)}P_\infty$ and
$T_{(\lambda x,e)}P_\infty$.
Using the same identif\/ication, from continuity (smoothness) of $\psi$ and $\psi_{0}=0$ we obtain
\begin{gather*}
0=\lim_{\lambda \rightarrow 0}\psi_{\lambda x}\big(\vec{0}_{\mathfrak{g}},\vec{w}\big)=\lim_{\lambda \rightarrow
0}\frac{1}{\lambda} \psi_{x}\big(\vec{0}_{\mathfrak{g}},\vec{w}\big)
\qquad\forall\, x\in \mathbb{R}^n,\qquad\forall\, \vec{w}\in T_{(x,e)}P_\infty
\end{gather*}
so that $\psi_{x}\big(\vec{0}_{\mathfrak{g}},\cdot\big)=0$ for all $x\in \mathbb{R}^n$, hence $\psi=0$
by~\eqref{eq:conda}.
Finally, it is straightforward to see that $(\Phi^{*}\omega_0)|_{\mathfrak{g}\times TP_\infty}=\psi=0$ holds.
\item
Let $P'=\mathbb{R}^n\backslash \{0\}\times S$ and $\Phi$ be def\/ined as above.
Then $K\times\{e\}$, for the unit-sphere $K:=\{x\in \mathbb{R}^n\:|\: \|x\|=1\}$, is a~$\Phi$-covering of $P'$ with the
properties from Example~\ref{example:SCHSV}.
Evaluating the corresponding conditions $i'')$, $ii'')$, $iii'')$, immediately shows that the set
of $\Phi$-invariant connections on $P'$ is in bijection with the smooth maps $\psi\colon \mathbb{R} \times
TK\rightarrow\mathfrak{s}$ for which $\psi|_{\mathbb{R} \times T_kK}$ is linear for all $k\in K$.
The corresponding invariant connections are given~by
\begin{gather*}
\omega^\psi_{(x,s)}(\vec{v}_x,\vec{\sigma}_s)
=\psi\left(\textstyle\frac{1}{\|x\|}\mathrm{pr}_{\|}(\vec{v}_x),\mathrm{pr}_{\perp}(\vec{v}_x)\right)
+s^{-1}\vec{\sigma}_s
\qquad\forall\, (\vec{v}_x,\vec{\sigma}_s)\in T_{(x,s)}P'.
\end{gather*}
Here, $\mathrm{pr}_{\|}$ denotes the projection onto the axis def\/ined by $x\in \mathbb{R}^n$, as well as $\mathrm{pr}_{\perp}$
the projection onto the corresponding orthogonal complement in $\mathbb{R}^n$.
\end{enumerate}
\end{Example}

 Also in the spherically symmetric case the $\varphi$-stabilizer of the origin has full dimension, and it turns
out to be convenient (cf.\ Appendix~\ref{subsec:IsotrConn}) to use the $\Phi$-covering $\mathbb{R}^3\times \{e\}$ in this situation as well.
Since the choice $P_\infty:=M\times\{e\}$ is always reasonable (cf.\ Lemma~\ref{lemma:mindimslice}.1) if there is a~point in the base manifold~$M$ (of the trivial bundle $M\times S$) whose
stabilizer is the whole group, we now adapt Theorem~\ref{th:InvConnes} to this situation.
To this end, we identify $T_xM$ with $T_{(x,e)}P_\infty$ for each $x\in M$ in the following.

\begin{scase}[trivial principal f\/ibre bundles]\label{scase:trivbundle}
Let $(G,\Phi)$ be a~Lie group of automorphisms of the trivial principal fibre bundle $P=M\times S$.
Then, the $\Phi$-invariant connections are in bijection with the smooth maps $\psi\colon \mathfrak{g}\times TM\rightarrow
\mathfrak{s}$ for which $\psi|_{\mathfrak{g}\times T_xM}$ is linear for all $x\in M$ and that fulfil the following
properties.

Let $\psi^{\pm}\left(\vec{g},\vec{v}_y,\vec{s}\right):=\psi\left(\vec{g},\vec{v}_y\right)\pm\vec{s}$ for
$((\vec{g},\vec{s}),\vec{v}_y)\in \mathfrak{q}\times T_yM$.
Then, for $q\in Q$, $x\in M$ with $q\cdot (x,e)=(y,e)\in M\times \{e\}$ and all $((\vec{g},\vec{s}),\vec{v}_x)\in
\mathfrak{q}\times T_xM$ we have
\begin{enumerate}\itemsep=0pt
\item[\rm i)] $\wt{g}(x,e) + \vec{v}_x -\vec{s}=0
\quad
\Longrightarrow
\quad
\psi^{-}(\vec{g},\vec{v}_x,\vec{s}\hspace{1pt})=0$,
\item[\rm ii)]
$\psi^+(\mathrm{d} L_q \vec{v}_x)=\rho(q)\circ
\psi\big(\vec{0}_{\mathfrak{g}},\vec{v}_x\big)\hspace*{11.4mm}\forall\, \vec{v}_x\in T_xM$,
\item[\rm iii)]
$\psi\big(\Add{q}(\vec{g}),\vec{0}_{y}\big)=\rho(q)\circ \psi\big(\vec{g},\vec{0}_{x}\big) \qquad\forall\, \vec{g}\in\mathfrak{g}$.
\end{enumerate}
\end{scase}

\begin{proof}
The elementary proof can be found in Appendix~\ref{sec:STPR}.
\end{proof}

\begin{Example}[spherically symmetric systems in loop quantum gravity]\label{bsp:Rotats}
Let $\varrho\colon {\rm SU}(2) \rightarrow {\rm SO}(3)$ be the universal covering map and $\sigma(x):=\varrho(\sigma)(x)$ for $x\in
\mathbb{R}^3$.
Moreover, let ${\mathfrak{z}}\colon \mathbb{R}^3 \rightarrow \mathfrak{su}(2)$ be def\/ined as in the second part of
Example~\ref{ex:eukl}.
We consider the action of $G={\rm SU}(2)$ on $P=\mathbb{R}^3\times {\rm SU}(2)$ def\/ined by $\Phi(\sigma,
(x,s)):=(\varrho(\sigma)(x),\sigma s)$.
It is shown in Appendix~\ref{subsec:IsotrConn} that the corresponding invariant connections are of the form
\begin{gather}
\begin{split}
& \omega^{abc}_{(x,s)}(\vec{v}_x,\vec{\sigma}_s):= \Add{s^{-1}}\big[a(x){\mathfrak{z}}(\vec{v}_x)+ b(x)[{\mathfrak{z}}(x),{\mathfrak{z}}(\vec{v}_x)]
\\
& \hphantom{\omega^{abc}_{(x,s)}(\vec{v}_x,\vec{\sigma}_s):=}{}
{}+c(x)[{\mathfrak{z}}(x),[{\mathfrak{z}}(x),{\mathfrak{z}}(\vec{v}_x)]]\big]+ s^{-1}\vec{\sigma}_s
\end{split} \label{eq:rotinvconn}
\end{gather}
for $(\vec{v}_x,\vec{\sigma}_s)\in T_{(x,s)}P$ and with rotation invariant maps $a,b,c\colon \mathbb{R}^3\rightarrow
\mathbb{R}$ for which the whole expression is a~smooth connection.

We claim that the functions $a$, $b$, $c$ can be assumed to be smooth as well.
More precisely, we show that we can assume that
\begin{gather*}
a(x)=f\big(\|x\|^2\big),
\qquad
b(x)=g\big(\|x\|^2\big),
\qquad
c(x)=h\big(\|x\|^2\big)
\end{gather*}
holds for smooth functions $f,g,h\colon (-\epsilon,\infty)\rightarrow \mathbb{R}$ with $\epsilon>0$.
Then, each pullback of such a~spherically symmetric connection by the global section $x\mapsto (x,e)$ can be written in
the form
\begin{gather*}
\wt{\omega}^{abc}_x(\vec{v}_x)=\widetilde{f}\big(\|x\|^2\big){\mathfrak{z}}(\vec{v}_x)+\widetilde{g}\big(\|x\|^2\big){\mathfrak{z}}(x\times \vec{v}_x)
+\widetilde{h}\big(\|x\|^2\big){\mathfrak{z}}\left(x\times (x\times \vec{v}_x)\right)
\end{gather*}
for smooth functions $\widetilde{f},\widetilde{g},\widetilde{h}\colon (-\epsilon,\infty)\rightarrow \mathbb{R}$ with $\epsilon>0$.

\begin{proof}[Proof of the claim]
1.~Smoothness of $\omega^{abc}$ implies smoothness of the real functions
\begin{gather*}
a_{\vec{n}}(\lambda):= a(\lambda\vec{n}),
\qquad
b_{\vec{n}}(\lambda):=\lambda b(\lambda\vec{n}),
\qquad
c_{\vec{n}}(\lambda):= \lambda^2 c(\lambda\vec{n})
\qquad\forall\, \lambda\in \mathbb{R}
\end{gather*}
for each $\vec{n}\in\mathbb{R}^3\backslash\{0\}$.
In fact, $a_{\vec{n}}(\lambda)\cdot {\mathfrak{z}}(\vec{n})=\omega^{abc}_{(\lambda \vec{n},e)}(\vec{n})$ is smooth, so that
smoothness of~$b_{\vec{n}}$ and~$c_{\vec{n}}$ is immediate from smoothness of $\lambda \mapsto \omega^{abc}_{(\lambda
\vec{e}_1,e)}(\vec{e}_2)$.

2.~Let $\vec{n}$ be f\/ixed.
Then, $a_{\vec{n}}$ is even so that $a_{\vec{n}}(\lambda){=}f (\lambda^2 )$ for a~smooth function $f\colon\!
(-\epsilon_1,\infty){\rightarrow} \mathbb{R}$, see~\cite{HasslerWhitneyb}.
Moreover, $b_{\vec{n}}$ is smooth and odd, so that $b_{\vec{n}}(\lambda) =\lambda g\big(\lambda^2\big)$ for a smooth
function $g\colon (-\epsilon_2,\infty)\rightarrow \mathbb{R}$, again by~\cite{HasslerWhitneyb}.
Similarly, $c_{\vec{n}}(\lambda)= l(\lambda^2)$ for a~smooth function $l\colon (-\epsilon_3,\infty)\rightarrow
\mathbb{R}$.
Since $\lambda \mapsto l(\lambda^2)$ is even and $l(0)=0$, for $s\in \mathbb{N}_{>0}$ Taylor's formula yields
\begin{gather*}
l\big(x^2\big)= a_1 x^2 +\dots + a_{s}x^{2s} + x^{2(s+1)}\phi(x)
=x^2 \big(a_1 +\dots + a_{s}x^{2s-2} + x^{2s}\phi(x)\big)=x^2L(x)
\end{gather*}
with remainder term $\phi(x):=\frac{1}{(2s+1)!}\frac{1}{x^{2s+2}}\int_{0}^{x}(x-t)\,l^{(2s+2)}(t)\,
\mathrm{d} t$ for $x\neq 0$ and $\phi(0):=l^{(2s+2)}(0)$.
Now, $\phi$ is continuous by Theorem 1 in~\cite{HasslerWhitneya}, so that $L$ is continuous as well.
But $x\mapsto x^2 L(x)$ is smooth, so that Corollary 1 in~\cite{HasslerWhitneya} shows that $L$ is smooth as well.
Now, $L$ is even, hence $L(x)=h(x^2)$ for some smooth function $h\colon (-\epsilon_4,\infty )\rightarrow \mathbb{R}$.
Then, $c_{\vec{n}}(\lambda)=l\big(\lambda^2\big)=\lambda^2h\big(\lambda^2\big)$, and for $x\neq 0$ we get
\begin{gather*}
b(x)=\|x\|\,b\left(\|x\|\frac{x}{\|x\|}\right)\frac{1}{\|x\|}= b_{\textstyle{\frac{x}{\|x\|}}}(\|x\|)\frac{1}{\|x\|}
=g\left(\|x\|^2\right),
\\
c(x)=\|x\|^2c\left(\|x\|\frac{x}{\|x\|}\right)\frac{1}{\|x\|^2}
=c_{\textstyle\frac{x}{\|x\|}}\left(\|x\|\right)\frac{1}{\|x\|^2} =h\left(\|x\|^2\right).
\end{gather*}
Moreover, for $x=0$ we have
\begin{gather*}
\hspace{31.5pt}b(x)[{\mathfrak{z}}(x),{\mathfrak{z}}(\vec{v}_x)]=0=g\big(\|x\|^2\big)[{\mathfrak{z}}(x),{\mathfrak{z}}(\vec{v}_x)],
\\
c(x)[{\mathfrak{z}}(x),[{\mathfrak{z}}(x),{\mathfrak{z}}(\vec{v}_x)]\big]=0=h(x)[{\mathfrak{z}}(x),[{\mathfrak{z}}(x),{\mathfrak{z}}(\vec{v}_x)]\big]
\end{gather*}
so that we can assume $a(x)=f(\|x\|^2)$, $b(x)=g(\|x\|^2)$ and $c(x)=h(\|x\|^2)$ for the smooth functions $f,g,h\colon
\left(-\min(\epsilon_1,\dots,\epsilon_4),\infty\right)\rightarrow \mathbb{R}$.
\end{proof}

In particular, there are spherically symmetric connections on $\mathbb{R}^3\backslash \{0\}\times {\rm SU}(2)$
which cannot be extended to those on~$P$.
For instance, if $b=c=0$ and $a(x):=1\slash \|x\|$ for $x\in \mathbb{R}^3\backslash \{0\}$, then~$\omega^{abc}$ cannot
be extended smoothly to an invariant connection on~$\mathbb{R}^3\times {\rm SU}(2)$ since elsewise~$a_{\vec{n}}$ could be
extended to a~continuous (smooth) function on $\mathbb{R}$.
\end{Example}

\section{Conclusions}

We conclude with a~short review of the particular cases that follow from
Theorem~\ref{th:InvConnes}.
For this let $(G,\Phi)$ be a~Lie group of automorphisms of the principal f\/ibre bundle $(P,\pi,M,S)$ and $\varphi$ the
induced action on~$M$.
\begin{itemize}\itemsep=0pt
\item
If $P=M\times S$ is trivial, then $M\times \{e\}$ is a~$\Phi$-covering of $P$.
As we have demonstrated in the spherically symmetric and scale invariant case (cf.\
Examples~\ref{ex:OnePoint} and~\ref{bsp:Rotats}), this choice can be useful for calculations if there is a~point in~$M$
whose $\varphi$-stabilizer is the whole group~$G$.
\item
If there is an element $x\in M$ which is contained in the closure of each $\varphi$-orbit,
 each $\Theta$-patch which contains some $p\in\pi^{-1}(x)$ is a~$\Phi$-covering of $P$, see Example~\ref{ex:Bruhat}.
If~$\varphi$ acts transitively on~$M$, for each $p\in P$ the zero-dimensional submanifold $\{p\}$ is
a~$\Phi$-covering of $P$; giving back Wang's original theorem, see Case~\ref{th:wang} and Example~\ref{ex:eukl}.
\item
Let $\Phi$ act via gauge transformations on $P$.
In this case each open covering $\{U_\alpha\}_{\alpha\in I}$ of~$M$ together with smooth sections $s_\alpha \colon
U_\alpha \rightarrow P$ provides the $\Phi$-covering $\{s_\alpha(U_\alpha)\}_{\alpha\in I}$ of $P$.
If~$G$ acts trivially, this specializes to the usual description of smooth connections by means of consistent families
of local 1-forms on the base manifold~$M$.
\item
If $P_0$ is a~$\Theta$-patch such that $\pi(P_0)$ intersects each $\varphi$-orbit in a~unique point,
it is a~$\Phi$-covering.
If in addition the stabilizer $Q_p$ does not depend on $p\in P_0$, we get back the characterization from~\cite{HarSni}, see
Example~\ref{example:SCHSV}.
\item
Assume there is a~collection of $\varphi$-orbits forming an open subset $U\subseteq M$.
Then, $O:=\pi^{-1}(U)$ is a~principal f\/ibre bundle and each $\Phi$-invariant connection on~$P$ restricts to
a~$\Phi$-invariant connection on~$O$.
Conversely, if~$U$ is in addition dense in~$M$, one can ask the question whether a~$\Phi$-invariant connection on~$O$ extends to a~$\Phi$-invariant connection on~$P$.
Since such an extension is necessarily unique (continuity), $\varphi$-orbits not contained in~$U$ can be seen as sources
of obstructions for the extendability of invariant connections on~$O$ to~$P$.
Indeed, as the examples in Subsection~\ref{sec:ApplTrivB} show, smoothness of these extension can give crucial
restrictions.
Moreover, by Example~\ref{ex:Bruhat}, taking one additional orbit into account can shrink the number of invariant
connections to zero.
Of particular interest, in this context, is the case where~$G$ is compact, as then the orbits of principal type always
form a~dense and open subset of~$M$ on which the situation of~\cite{HarSni} always holds locally~\cite{RumSchmBuch}. This gives rise to
a~canonical $\Phi$-covering $O$ consisting of convenient patches.
Thus, using the present characterization theorem, there is a~realistic chance to get some general classif\/ication results in
the compact case\footnote{To be used, e.g., to extend the framework of the foundational LQG reduction
paper~\cite{BojoKa}.}.
\end{itemize}

As Corollary~\ref{cor:reductions} shows, in the general situation one can always construct $\Phi$-coverings of $P$ from families of
$\varphi$-patches in~$M$.
In particular, the f\/irst three cases arise in this way.

\appendix

\section*{Appendix}

\section{A technical proof}\label{sec:STPR}

%\subsection{The proof of Case~\ref{scase:trivbundle}}\label{subsec:TrivBund}

\begin{proof}[Proof of Case~\ref{scase:trivbundle}]
The only patch is $M\times\{e\}$, so that a~reduced connection is a~smooth
map $\psi\colon \mathfrak{g}\times TM \rightarrow \mathfrak{s}$ with the claimed linearity property
and that fulf\/ils the
two conditions from Corollary~\ref{cor:psialpha}.
Obviously, $ii)$ and~iii) are equivalent.
Moreover, i) follows from $i)$ for $p_\alpha=p_\beta=(x,e)$, $q=(e,e)$, $\vec{w}_{p_\beta}=\vec{v}_x$
and $\vec{w}_{p_\alpha}=\vec{0}_{(x,e)}$, see also $a)$ in Remark~\ref{rem:Psialphaconderkl}.
Now, to obtain ii), let $\vec{v}_x\in T_xM$, $q\in Q$ and $q\cdot (x,e)=(y,e)$.
Then, $\mathrm{d} L_q\vec{v}_x=(\vec{v}_y,-\vec{s})$ for elements $\vec{v}_y\in T_yM$ and $\vec{s}\in \mathfrak{s}$ so
that
\begin{gather*}
\psi^+(\mathrm{d}
L_q\vec{v}_x)=\psi^+(\vec{v}_y,-\vec{s}\hspace{1pt})=\psi\big(\vec{0}_{\mathfrak{g}},\vec{v}_y\big)-\vec{s}
\stackrel{i)}{=}\rho(q)\circ\psi\big(\vec{0}_{\mathfrak{g}},\vec{v}_x\big).
\end{gather*}
It remains to show that~i) and~ii) imply $i)$.
To this end, let $(y,e)=q\cdot (x,e)$ for $x,y\in M$ and $q\in Q$.
Then $i)$ reads
\begin{gather*}
\wt{g}(y,e)+ \vec{v}_{y}-\vec{s}=\mathrm{d} L_q\vec{v}_{x}
\quad
\Longrightarrow
\quad
\psi^-(\vec{g},\vec{v}_{y},\vec{s}\hspace{1pt})=\rho(q)\circ\psi\big(\vec{0}_{\mathfrak{g}},\vec{v}_{x}\big),
\end{gather*}
where $\vec{v}_{x}\in T_xM$, $\vec{v}_{y}\in T_yM$, $\vec{s}\in \mathfrak{s}$ and $\vec{g}\in \mathfrak{g}$.
Let $\mathrm{d} L_q\vec{v}_x=(\vec{v}_y,-\vec{s}\hspace{1pt})$ be as above.
If~ii) is true, then it is clear from
\begin{gather*}
\psi^{-}(\vec{v}_y,\vec{s}\hspace{1pt})=\psi^+(\mathrm{d} L_q\vec{v}_x)\stackrel{\rm ii)}{=} \rho(q)\circ
\psi\big(\vec{0}_{\mathfrak{g}},\vec{v}_x\big)
\end{gather*}
that $i)$ is true for $\big(\big(\vec{0}_{\mathfrak{g}},\vec{s}\hspace{1pt}),\vec{v}_y\big)$, i.e.,
\begin{gather*}
\vec{0}_{\mathfrak{g}}+\vec{v}_y-\vec{s}=\mathrm{d} L_q\vec{v}_{x}
\quad
\Longrightarrow
\quad
\psi\big(\vec{0}_{\mathfrak{g}},\vec{v}_y\big)-\vec{s}=\rho(q)\circ \psi\big(\vec{0}_{\mathfrak{g}},\vec{v}_x\big).
\end{gather*}
Due to~i) and the linearity properties of $\psi$, the condition $i)$ then also holds for each other
element $((\vec{g}\hspace{1pt}{}',\vec{s}\hspace{1pt}{}'),\vec{v}\hspace{1pt}{}'_y)\in \mathfrak{q}\times T_yM$ with
$\wt{g}\hspace{1pt}{}'(y,e)+ \vec{v}\hspace{1pt}{}'_{y}-\vec{s}\hspace{1pt}{}'=\mathrm{d} L_q\vec{v}_{x}$.
\end{proof}

\section{Technical calculations}%\label{sec:AppTechCalc}

Let $P=\mathbb{R}^3\times {\rm SU}(2)$, $\varrho \colon {\rm SU}(2)\rightarrow {\rm SO}(3)$ the universal covering map, $E=\mathbb{R}^3
\rtimes_\varrho {\rm SU}(2)$ and ${\mathfrak{z}}\colon \mathbb{R}^3\rightarrow \mathfrak{su}(2)$ be def\/ined as in the second part of
Example~\ref{ex:eukl}.
Then, $\varrho(\sigma)={\mathfrak{z}}^{-1}\circ\Add{\sigma}\circ {\mathfrak{z}} $ and each $\sigma\in {\rm SU}(2)$ can be written as
\begin{gather*}
\sigma= \cos(\alpha/2)\mathds{1} + \sin(\alpha/2){\mathfrak{z}}(\vec{n})=\exp\big(\alpha/2\cdot{\mathfrak{z}}(\vec{n})\big)
\end{gather*}
for some $|\vec{n}|=1$ and $\alpha\in [0,2\pi]$.
In this case $\varrho(\sigma)$ rotates a~point $x$ by the angle $\alpha$ w.r.t.\ the axis~$\vec{n}$.
For simplicity, if $\sigma\in {\rm SU}(2)$ and $x\in \mathbb{R}^3$, we write $\sigma(x)$ instead of $\varrho(\sigma)(x)$ in
the following.

\subsection{A result used in the end of Section~\ref{sec:PartCases}}\label{subsec:DifferentLifts}

We consider the f\/ibre transitive action $\Phi'\colon E\times P\rightarrow P$ def\/ined by
$\Phi'((v,\sigma),(x,s)):=(v+ \sigma(x),s)$ and claim that the connection
\begin{gather*}
\omega_0(\vec{v}_x,\vec{\sigma}_s)=s^{-1}\vec{\sigma}_s
\qquad\forall\, (\vec{v}_x,\vec{\sigma}_s) \in T_{(x,s)}P
\end{gather*}
is the only $\Phi'$-invariant one.
For this, observe that the stabilizer of $x=0$ w.r.t.~$\varphi'$ is given by ${\rm SU}(2)$ and $\phi'_{(0,e)}(\sigma)=e$ for
all $\sigma\in {\rm SU}(2)$.
We apply Wang's theorem to $p=(0,e)$.
Then condition~a) yields $\psi(\vec{s})=0$ for all $\vec{s}\in \mathfrak{su}(2)$, and~b)
now reads $\psi\circ \Add{\sigma}=\psi$ for all $\sigma\in {\rm SU}(2)$.
Consequently, for $\vec{v}\in \mathbb{R}^3\subseteq \mathfrak{e}=\mathbb{R}^3 \times \mathfrak{su}(2)$ we obtain
\begin{gather*}
0=\dttB{t}{0}\,\psi(\vec{v}\hspace{1pt})=\dttB{t}{0}\, \psi\circ \Add{\exp(t\vec{s})}(\vec{v}\hspace{1pt})
=\psi\left(\dttB{t}{0}\,\varrho(\exp(t\vec{s}\hspace{1pt}))(\vec{v}\hspace{1pt})\right)=\psi \circ {\mathfrak{z}}^{-1}([\vec{s},{\mathfrak{z}}(\vec{v}\hspace{1pt})])
\end{gather*}
for all $\vec{s}\in \mathfrak{su}(2)$, just by linearity of $\psi$.
This gives
\begin{gather*}
0=\psi\big({\mathfrak{z}}^{-1}([\tau_i,{\mathfrak{z}}(\vec{e}_j)])\big)=\psi\big({\mathfrak{z}}^{-1}([\tau_i,\tau_j])\big)=2
\epsilon_{ijk}\psi(\vec{e}_k),
\end{gather*}
hence $\psi=0=\Phi^{\prime*}_{p}\omega_0$.

\subsection{Spherically symmetric connections}\label{subsec:IsotrConn}

We consider the action $\Phi$ of ${\rm SU}(2)$ on $P$ def\/ined by $\Phi(\sigma, (x,s)):=(\sigma(x),\sigma s)$, and show that the
corresponding invariant connections are given by (see~\eqref{eq:rotinvconn} in Example~\ref{bsp:Rotats})
\begin{gather*}
\omega^{abc}_{(x,s)}(\vec{v}_x,\vec{\sigma}_s):= \Add{s^{-1}}\big[a(x){\mathfrak{z}}(\vec{v}_x)+ b(x)[{\mathfrak{z}}(x),{\mathfrak{z}}(\vec{v}_x)]
+c(x)[{\mathfrak{z}}(x),[{\mathfrak{z}}(x),{\mathfrak{z}}(\vec{v}_x)]]\big]+ s^{-1}\vec{\sigma}_s
\end{gather*}
with rotation invariant maps $a,b,c\colon \mathbb{R}^3\rightarrow \mathbb{R}$ for which the whole expression is a~smooth
connection.
Now, a~straightforward calculation shows that each $\omega^{abc}$ is $\Phi$-invariant, so that it remains to verify that
each $\Phi$-invariant connection is of the upper form.
For this, we reduce the connections $\omega^{abc}$ w.r.t.~$P_\infty=\mathbb{R}^3\times \{e\}$ and show that each map
$\psi$ as in Case~\ref{scase:trivbundle} can be obtained in this way.
To this end, let $\vec{g}\in \mathfrak{g}$, $p=(x,e)\in P_\infty$ and $\gamma_x\colon (-\epsilon,\epsilon)\rightarrow M$ be
a~smooth curve with $\dot\gamma_x(0)=\vec{v}_x\in T_xM\subseteq T_pP_\infty$.
Then,
\begin{gather}
\begin{split}
& \mathrm{d}_{(e,p)}\Phi(\vec{g},\vec{v}_x)
=\left(\dttB{t}{0}{\mathfrak{z}}^{-1}\left(\exp(t\vec{g}\hspace{1pt}){\mathfrak{z}}(\gamma_x(t))\exp(t\vec{g}\hspace{1pt})^{-1}\right),
\exp(t\vec{g}\hspace{1pt})\right)
\\
& \phantom{\mathrm{d}_{(e,p)}\Phi(\vec{g},\vec{v}_x)}
=\left({\mathfrak{z}}^{-1}\left([\vec{g},{\mathfrak{z}}(x)]\right)+\vec{v}_x,\vec{g}\hspace{1pt}\right).
\end{split}
\label{eq:theta}
\end{gather}
This equals $\vec{s}$ if\/f $\vec{g}=\vec{s}$ and $\vec{v}_x={\mathfrak{z}}^{-1}([{\mathfrak{z}}(x),\vec{g}\hspace{1pt}])$.
Consequently, for the reduced connection $\psi^{abc}$ which corresponds to $\omega^{abc}$ we obtain
\begin{gather*}
\begin{split}
& \psi^{abc}(\vec{g},\vec{v}_x) =\big(\Phi^{*}\omega^{abc}\big)_{(e,p)}(\vec{g},\vec{v}_x)
=\omega^{abc}_{p}\left({\mathfrak{z}}^{-1}\left([\vec{g},{\mathfrak{z}}(x)]+{\mathfrak{z}}(\vec{v}_x)\right),\vec{g}\hspace{1pt}\right)
\\
& \phantom{\psi^{abc}(\vec{g},\vec{v}_x)}
=a(x)\big[[\vec{g},{\mathfrak{z}}(x)]+{\mathfrak{z}}(\vec{v}_x)\big]+ b(x)\big[[{\mathfrak{z}}(x),[\vec{g},{\mathfrak{z}}(x)]]+[{\mathfrak{z}}(x),{\mathfrak{z}}(\vec{v}_x)]\big]
\\
&\phantom{\psi^{abc}(\vec{g},\vec{v}_x)=}
{}+c(x)\big[[{\mathfrak{z}}(x),[{\mathfrak{z}}(x),[\vec{g},{\mathfrak{z}}(x)]]]+[{\mathfrak{z}}(x),[{\mathfrak{z}}(x),{\mathfrak{z}}(\vec{v}_x)]]\big] +\vec{g}.
\end{split}
\end{gather*}
Now, assume that $\psi$ is as in Case~\ref{scase:trivbundle}.
Then for $q\in Q$ and $p\in P_\infty$ we have $q\cdot p\in P_\infty$ if\/f $q=(\sigma,\sigma)$ for some $\sigma\in {\rm SU}(2)$
and $p=(x,e)$ for some $x\in M$.
Consequently, $q\cdot p=(\sigma(x),e)$ as well as $\mathrm{d} L_q(\vec{v}_x)=\sigma(\vec{v}_x)$ for all $\vec{v}_x\in
T_xM$ so that~ii) gives
\begin{gather*}
\psi\big(\vec{0}_{\mathfrak{g}},\sigma(\vec{v}_x)\big)=\psi^+(\mathrm{d} L_q(\vec{v}_x))=\Add{\sigma}\circ
\psi\big(\vec{0}_{\mathfrak{g}},\vec{v}_x\big),
\end{gather*}
hence
\begin{gather}
\label{eq:condrei}
\psi\big(\vec{0}_{\mathfrak{g}},\vec{v}_x\big)=\Add{\sigma^{-1}}\circ
\psi\big(\vec{0}_{\mathfrak{g}},\sigma(\vec{v}_x)\big)
\qquad\forall\, \vec{v}_x\in T_xM.
\end{gather}
If $x\neq 0$, then for $\sigma_t:=\exp(t{\mathfrak{z}}(x))$ we have $\sigma_t(x)=x$ and $\sigma_t(\vec{v}_x)\in T_x M$
for all $t\in \mathbb{R}$.
Then, linearity of $\psi_x:=\psi|_{\mathfrak{g}\times T_{(x,e)}P_\infty}$ yields \begin{gather*}
0=\dttB{t}{0}\psi\big(\vec{0}_{\mathfrak{g}},\vec{v}_x\big)
\stackrel{\eqref{eq:condrei}}{=}\dttB{t}{0}\Add{\sigma_t^{-1}}\circ
\psi\big(\vec{0}_{\mathfrak{g}},\sigma_t(\vec{v}_x)\big)
\\
\phantom{0}=\dttB{t}{0}\sigma^{-1}_t \big(\psi_x\circ {\mathfrak{z}}^{-1}\big)\left(\sigma_t\,{\mathfrak{z}}(\vec{v}_x)\,\sigma^{-1}_t\right)
\sigma_t
\\
\phantom{0}\hspace*{-0.5mm}\stackrel{\text{lin.}}{=}-{\mathfrak{z}}(x) \psi_x\big(\vec{0}_{\mathfrak{g}},\vec{v}_x\big)+ \big(\psi_x\circ
{\mathfrak{z}}^{-1}\big)\left[{\mathfrak{z}}(x){\mathfrak{z}}(\vec{v}_x)-{\mathfrak{z}}(\vec{v}_x){\mathfrak{z}}(x)\right]
+\psi_x\big(\vec{0}_{\mathfrak{g}},\vec{v}_x\big){\mathfrak{z}}(x),
\end{gather*}
hence $\big[{\mathfrak{z}}(x),\psi\big(\vec{0}_{\mathfrak{g}},\vec{v}_x\big)\big]=\big(\psi\circ
{\mathfrak{z}}^{-1}\big)\left([{\mathfrak{z}}(x),{\mathfrak{z}}(\vec{v}_x)]\right)$.
For $x=\lambda \vec{e}_1\neq 0$ and $\kappa_j:=\psi\big(\vec{0}_{\mathfrak{g}},\vec{v}_x\big)$ with
$\vec{v}_x=\vec{e}_j$ this reads
\begin{gather*}
\left[\tau_1,\kappa_j
\right]=\big(\psi_x\circ{\mathfrak{z}}^{-1}\big)([\tau_1,\tau_j])=\big(\psi_x\circ{\mathfrak{z}}^{-1}\big)(2\epsilon_{1jk}\tau_k)
=2\epsilon_{1jk}\psi_x\big(\vec{0}_{\mathfrak{g}},\vec{e}_k\big)=2\epsilon_{1jk}\kappa_k.
\end{gather*}
From these relations, it follows that
\begin{gather*}
\kappa_1=r(\lambda)\tau_1,
\qquad
\kappa_2=s(\lambda)\tau_2+ t(\lambda)\tau_3,
\qquad
\kappa_3=s(\lambda)\tau_3- t(\lambda)\tau_2
\end{gather*}
for real constants $r(\lambda),s(\lambda),t(\lambda)$ depending on $\lambda\in\mathbb{R}\backslash\{0\}$.
Then, for $x=\lambda \vec{e}_1$ and
\begin{gather*}
a(\lambda\vec{e}_1):=r(\lambda),
\qquad
b(\lambda \vec{e}_1):=\frac{t(\lambda)}{2\lambda},
\qquad
c(\lambda\vec{e}_1):=\frac{r(\lambda)-s(\lambda)}{4\lambda^2}
\end{gather*}
linearity of $\psi_x$ yields
\begin{gather*}
\psi\big(\vec{0}_{\mathfrak{g}},\vec{v}_x\big)=a(x) {\mathfrak{z}}(\vec{v}_x)+
b(x)[{\mathfrak{z}}(x),{\mathfrak{z}}(\vec{v}_x)]+c(x)[{\mathfrak{z}}(x),[{\mathfrak{z}}(x),{\mathfrak{z}}(\vec{v}_x)]].
\end{gather*}
Now, if $x\neq 0$ is arbitrary, then $x=\sigma(\lambda\vec{e}_1)$ for some $\sigma\in {\rm SU}(2)$ and $\lambda > 0$.
So, $(\sigma,\sigma)\cdot (\lambda \vec{e}_1,e)=(x,e)$ and if we consider $\sigma^{-1}(\vec{v}_x)$ as an element of
$T_{(\lambda\vec{e_1},e)}P_\infty$, then~ii) gives
\begin{gather*}
\psi\big(\vec{0}_{\mathfrak{g}},\vec{v}_x\big)
=\psi^+(\vec{v}_x) =\psi^+\big(\mathrm{d}L_{(\sigma,\sigma)}\big(\sigma^{-1}(\vec{v}_x)\big)\big)
\\
\phantom{\psi\big(\vec{0}_{\mathfrak{g}},\vec{v}_x\big)}
\stackrel{\rm ii)}{=}\Add{\sigma}\circ
\psi^+\big(\sigma^{-1}(\vec{v}_x)\big)=\Add{\sigma}\circ \psi\big(\vec{0}_{\mathfrak{g}},\sigma^{-1}(\vec{v}_x)\big)
\\
\phantom{\psi\big(\vec{0}_{\mathfrak{g}},\vec{v}_x\big)}
=a(\lambda\vec{e}_1){\mathfrak{z}}(\vec{v}_x)+b(\lambda\vec{e}_1)\left[{\mathfrak{z}}(x),{\mathfrak{z}}(\vec{v}_x)\right]+c(\lambda\vec{e}_1)
[{\mathfrak{z}}(x),[{\mathfrak{z}}(x),{\mathfrak{z}}(\vec{v}_x)]].
\end{gather*}
\noindent For $x=0$ we have $\sigma(x)=x$ for all $\sigma\in {\rm SU}(2)$, and analogous to the case $x\neq 0$, but now for
$\sigma_t:=\exp(t\vec{g})$ with $\vec{g}\in \mathfrak{g}$, we obtain from~\eqref{eq:condrei} that
\begin{gather*}
\big[\vec{g},\psi_0\big(\vec{0}_{\mathfrak{g}},\vec{v}_0\big)\big]=\big(\psi_0\circ
{\mathfrak{z}}^{-1}\big)\left([\vec{g},{\mathfrak{z}}(\vec{v}_0)]\right)
\qquad\forall\, \vec{g}\in\mathfrak{su}(2), \qquad\forall\, \vec{v}_0\in T_0M.
\end{gather*}
This gives
$\big[\tau_i,\psi\big(\vec{0}_{\mathfrak{g}},\vec{e}_j\big)\big]=2\epsilon_{ijk}\psi\big(\vec{0}_{\mathfrak{g}},\vec{e}_k\big)$
and forces $\psi(\vec{v}_0)=a(0){\mathfrak{z}}(\vec{v}_0)$ for all $\vec{v}_0\in T_{(0,e)}P_\infty$ whereby $a(0)\in \mathbb{R}$ is
some constant.
Together, this shows
\begin{gather*}
\psi\big(\vec{0}_{\mathfrak{g}},\vec{v}_x\big)=a(x){\mathfrak{z}}(\vec{v}_x)+b(x)[{\mathfrak{z}}(x),{\mathfrak{z}}(\vec{v}_x)]+c(x)[{\mathfrak{z}}(x),[{\mathfrak{z}}(x),{\mathfrak{z}}(\vec{v}_x)]]
\end{gather*}
with functions $a$, $b$, $c$ that depend on $\|x\|$ in such a~way that the whole expression is smooth.
Finally, to determine $\psi\big(\vec{g},\vec{0}_x\big)$ for $\vec{g}\in \mathfrak{su}(2)=\mathfrak{g}$, we consider
${\mathfrak{z}}^{-1}([{\mathfrak{z}}(x),\vec{g}])$ as an element of~$T_{(x,e)}P_\infty$.
Then by~\eqref{eq:theta} we obtain from~i) that $\psi\big(\vec{g},{\mathfrak{z}}^{-1}([{\mathfrak{z}}(x),\vec{g}])\big)-\vec{g}=0$,
hence
\begin{gather*}
\psi\big(\vec{g},\vec{0}_x\big)=\vec{g}-\psi\big(\vec{0}_{\mathfrak{g}},{\mathfrak{z}}^{-1}([{\mathfrak{z}}(x),\vec{g}\hspace{1pt}])\big)
\\
\phantom{\psi\big(\vec{g},\vec{0}_x\big)}
=\vec{g}-a(x)[{\mathfrak{z}}(x),\vec{g}\hspace{1pt}]-b(x)[{\mathfrak{z}}(x),[{\mathfrak{z}}(x),\vec{g}\hspace{1pt}]]-c(x)[{\mathfrak{z}}(x),[{\mathfrak{z}}(x),[{\mathfrak{z}}(x),\vec{g}\hspace{1pt}]]]
\\
\phantom{\psi\big(\vec{g},\vec{0}_x\big)}
=a(x)[\vec{g},{\mathfrak{z}}(x)]+b(x)[{\mathfrak{z}}(x),[\vec{g},{\mathfrak{z}}(x)]]+c(x)[{\mathfrak{z}}(x),[{\mathfrak{z}}(x),[{\mathfrak{z}}(x),\vec{g}\hspace{1pt}]]]+\vec{g}.
\end{gather*}

\subsection*{Acknowledgements} The author is grateful to the anonymous referees for several helpful comments and
suggestions.
Moreover, he thanks Christian Fleischhack for numerous discussions and many helpful comments on a~draft of the present
article.
He is grateful for discussion with various members of the math faculty of the University of Paderborn.
In particular, with Joachim Hilgert, Bernhard Kr\"{o}tz, Benjamin Schwarz and Andreas Schmied.
He also thanks Gerd Rudolph for general discussions and comments on a~f\/irst draft of this article.
The author has been supported by the Emmy-Noether-Programm of the Deutsche Forschungsgemeinschaft under grant
FL~622/1-1.

\pdfbookmark[1]{References}{ref}
\LastPageEnding

\end{document}